\documentclass[aps,pra,preprint,superscriptaddress,longbibliography]{revtex4-2}

\usepackage{amsmath}
\usepackage{amsthm}
\usepackage{amsfonts}
\usepackage{amssymb}
\usepackage{xcolor,graphicx}
\usepackage{tikz}
\usepackage{color}
\usepackage[pdftex]{hyperref} 
\usepackage{longtable}
\usepackage{array}
\usepackage{dsfont}

\begin{document}
	
	\title{Optical coupling control of isolated mechanical resonators}
	
	\author{F.~E. Onah}
	\email[e-mail: ]{a00834081@tec.mx}
	\affiliation{Tecnol\'ogico de Monterrey, Escuela de Ingenier\'ia y Ciencias, Ave. Eugenio Garza Sada 2501, Monterrey, N.L., Mexico, 64849}
	\affiliation{The Division of Theoretical Physics, Physics and Astronomy, University of Nigeria Nsukka, Nsukka Campus, Enugu State, Nigeria}

	\author{B.~R. Jaramillo-\'Avila}
	\email[e-mail: ]{bjaramillo@cicese.mx}
	\affiliation{CONACYT - CICESE, Unidad Monterrey, Alianza Centro 504, PIIT Apodaca, Nuevo Leon, Mexico, 66629}
	
	\author{F.~H. Maldonado-Villamizar}
	\email[e-mail: ]{fmaldonado@inaoep.mx}
	\affiliation{CONACYT - Instituto Nacional de Astrof\'isica, \'Optica y Electr\'onica, Calle Luis Enrique Erro No. 1. Sta. Ma. Tonantzintla, Pue. C.P. 72840, Mexico}

	\author{B.~M. Rodr\'iguez-Lara}
	\email[e-mail: ]{bmlara@tec.mx}
	\affiliation{Tecnol\'ogico de Monterrey, Escuela de Ingenier\'ia y Ciencias, Ave. Eugenio Garza Sada 2501, Monterrey, N.L., Mexico, 64849}
	
\date{\today}
	
\begin{abstract}
    We present a Hamiltonian model describing two pairs of mechanical and optical modes under standard optomechanical interaction.
    The vibrational modes are mechanically isolated from each other and the optical modes couple evanescently.
    We recover the ranges for variables of interest, such as mechanical and optical resonant frequencies and naked coupling strengths, using a finite element model for a standard experimental realization.
    We show that the quantum model, under this parameter range and external optical driving, may be approximated into parametric interaction models for all involved modes. 
    As an example, we study the effect of detuning in the optical resonant frequencies modes and optical driving resolved to mechanical sidebands and show an optical beam splitter with interaction strength dressed by the mechanical excitation number, a mechanical bidirectional coupler, and a two-mode mechanical squeezer where the optical state mediates the interaction strength between the mechanical modes.
\end{abstract}

	\maketitle
	\newpage
	
\section{Introduction}
Optomechanical systems provide a versatile platform for quantum optics experiments and applications, including optical
bi-stability \cite{Dorsel1983, Aldana2013}, damping and anti-damping of mechanical motion in microwave-coupled mechanical resonators \cite{Cuthbertson1996,Aspelmeyer2014}, optically-assisted cooling of mechanical oscillations \cite{Mancini1998, Marquardt2007, Marquardt2008, Liu2013}, and  optomechanically induced transparency \cite{Weis2010, Karuza2013}, for example. 
They are a promising platform \cite{Frank2010, Qiao2018, Zhou2014, Yang2020, Pietikainen2022, Gu2018} to build sensors \cite{Qiao2018, Li2021} and quantum information transducers \cite{Guha2021, Balram2022} relying on the effect of electromagnetic radiation pressure on the vibrational modes of mechanical objects \cite{Braginsky1967a,  Braginsky1970}; for example, suspended micromirrors, membranes, microtoroids, microsphere resonators, micromembranes in superconducting circuits, 2D photonic crystals, photonic crystal nanobeams, and cold atoms in optical cavities \cite{Aspelmeyer2014}. Additionally, some of these platforms allow for further coupling between two or more optomechanical cavities, increasing the number of plausible applications for these systems \cite{Xu2015}. 

Recent advances in optomechanical cooling provide access to both mechanical and optical ground states and open the door to a wider range of low excitation number experiments \cite{Qiu2020}.
Optomechanical systems in the quantum regime may find use in quantum technologies.
For example, in quantum sensing and metrology, controlling the interaction of mechanical oscillators may lead to the engineering of two-mode squeezed states \cite{Xue2007, Tan2013, Woolley2014, Pontin2016, Mahbood2014, Patil2015, Shakeri2016}, or the development of mechanical couplers \cite{DeMartini2005, Piergentili2021} needed for mechanical interferometers.
In quantum information platforms, they may serve as transducers from microwave to optical spectrum \cite{Stannigel2010, Stannigel2011} or  mechanical memories \cite{Cole2011, Fiaschi2021}.

We are interested in the quantum dynamical description of two mechanically isolated vibrational modes, each one interacting with its own optical mode under standard optomechanical coupling. 
We introduce evanescent coupling between optical modes that allows for optical control of mechanical coupling under optical sideband driving.  
We present a finite element modeling analysis of plausible physical realizations for our model in Sec. \ref{sec:S2} in order to recover parameter ranges that may inform our analysis of the dynamics.
In Sec. \ref{sec:S3}, we introduce the quantum mechanical model and show that it is possible to define a reference frame where it takes the form of a parametric Hamiltonian where all mechanical and optical modes interact. 
In this reference frame, it becomes straightforward to realize that it is possible to induce and control the interaction of the mechanical modes by external optical sideband driving.
Then, we explore on-resonance driving of identically fabricated optical cavities and show that the effective model is that of an optical beam splitter where the coupling strength is modified by the state of the vibrational modes in Sec. \ref{sec:S4}.
We show that red sideband driving of the optical cavities with detuning equal to the mechanical frequency produces different effects depending on the detuning between the optical cavities. 
If the resonant frequency detunning between the optical cavities is equal to the difference between the mechanical resonant frequencies, optically mediated mechanical mode coupling appears, Sec \ref{sec:S5}. 
If it is equal to the sum of the mechanical resonant frequencies, optical mediated parametric mechanical coupling appears, Sec \ref{sec:S6}. 
In both cases, the optical state affects the coupling strength between the mechanical vibrational modes, we numerically explore these dynamics.
We close with our conclusion in Sec. \ref{sec:S7}.

\section{Finite Element Model} \label{sec:S2}
We are interested in a standard experimental optomechanical setup; a silica nanobeam with an engraved one-dimensional photonic defect cavity \cite{Eichenfield2009,Chan2011,Chan2012}. 
For the sake of simplicity, we consider a periodic array composed by 75 rectangular cells where a quadratic reduction in size for the middle 15 cells introduces a defect \cite{Eichenfield2009}.
We take each regular cell with length $360~\textrm{nm}$ ($x$-axis), width $1400~\textrm{nm}$ ($y$-axis), and thickness $220~\textrm{nm}$ ($z$-axis) and use finite element modeling (FEM) to find the principal optical and mechanical modes at $(2 \pi)\, 204~\mathrm{rad/sec}$ and $(2 \pi)\,2.23~\mathrm{rad/sec}$, in that order, in good agreement with experimental results \cite{Eichenfield2009}. 
Radiation pressure may induce a mechanical deformation that modifies the geometry of each optical cavity and, in consequence, its characteristic frequency, leading to optomechanical coupling.
Photonic crystal nanobeams of these scales lead to bare optomechanical couplings of the order of $g \sim (2 \pi)\, 10^{6}~\textrm{rad/sec}$ \cite{Chan2012}. 
These devices need to be pumped with an external laser whose power may vary from a few to hundreds of thousands of nanowatts, see supplementary material in Ref. \cite{Chan2012}, leading to laser-to-cavity coupling rates of the order of tens of $10^{6} ~\mathrm{rad/sec}$ \cite{Chan2012} and to pump rates between $\Omega_{\text{min}} \sim (2 \pi)\, 10^{9}~\textrm{rad/sec}$ and  $\Omega_{\text{max}} \sim (2 \pi)\, 10^{11}~\textrm{rad/sec}$.

\begin{figure}
\includegraphics{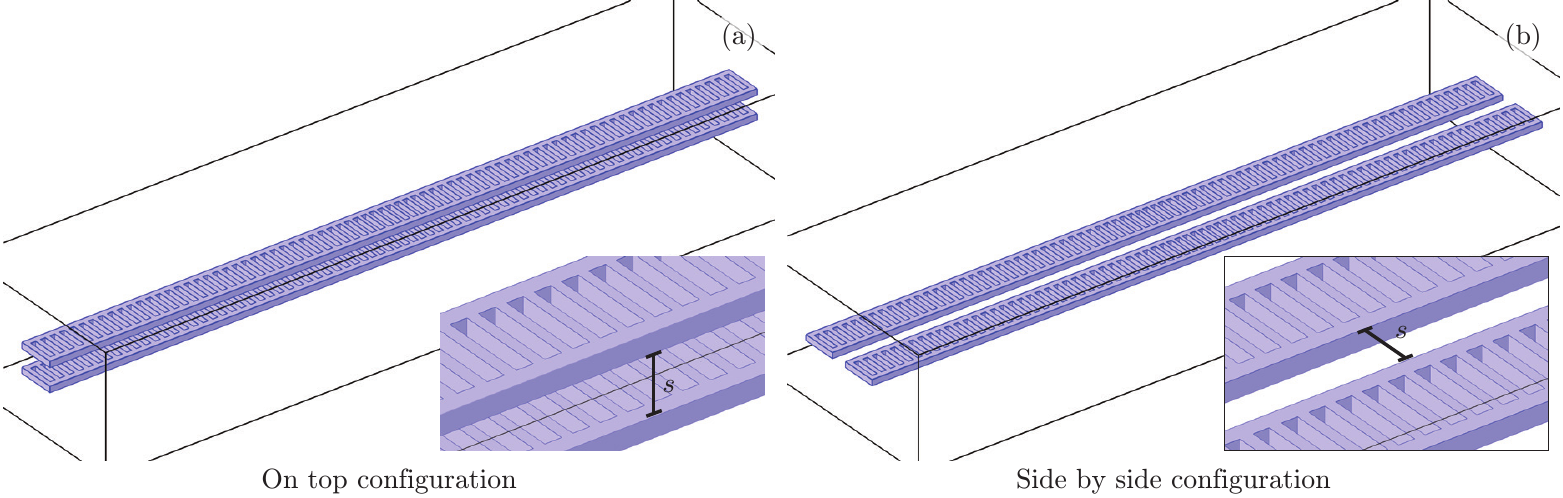}
\caption{(a) On-top and (b) side-by-side configurations coupling optical modes in two isolated optomechanical photonic crystal nanobeams.}\label{fig:1}
\end{figure}

\begin{figure}
\includegraphics{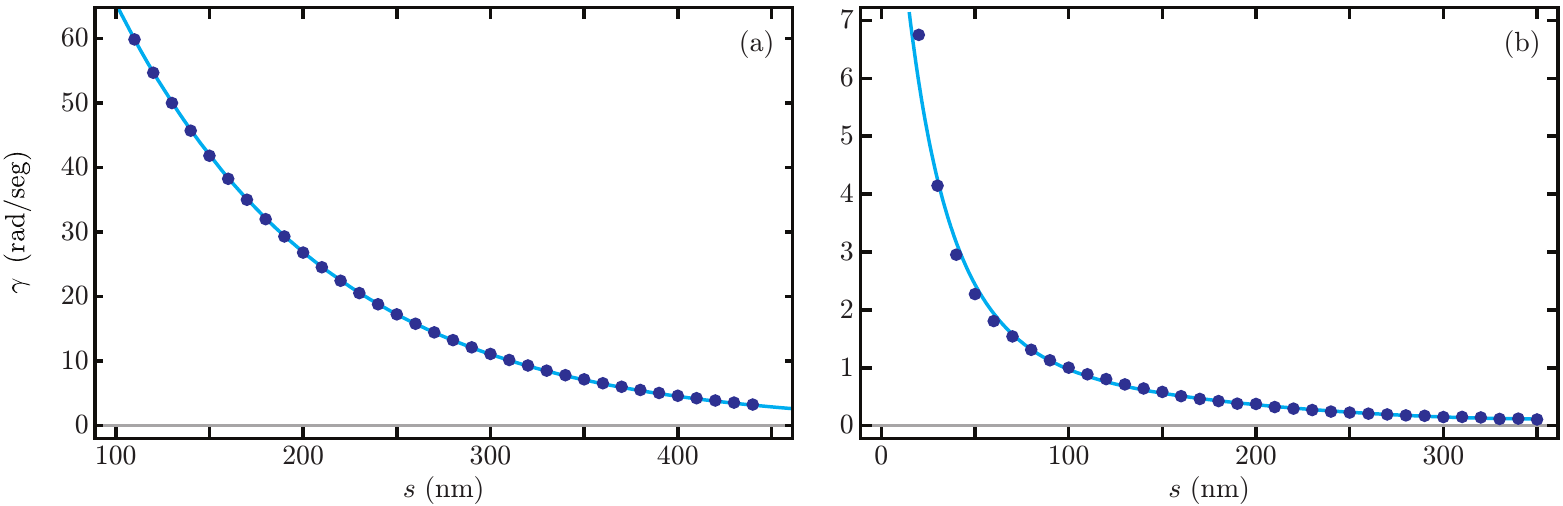}
\caption{Coupling between the fundamental optical modes in two identical optomechanical photonic crystal nanobeams as a function of the gap between the nanobeams $s$ for the (a) on-top and (b) side-by-side configurations. Dots show results from FEM and solid curves fit a simple exponential decay for the on-top configuration and fourth-order polynomial exponential decay for the side-by-side configuration.}\label{fig:2}
\end{figure}

In order to explore theoretically the optical coupling between two of these structures, we place two identical nanobeams parallel to each other and vary their separation. 
We use two possible configurations, one nanobeam on top of the other, Fig.~\ref{fig:1}(a) and two nanobeams side-by-side, Fig.~\ref{fig:1}(b). 
In both configurations, we consider the mechanical modes of each nanobeam isolated. 
The optical modes localized in each photonic defect cavity have evanescent fields outside its structure. 
These fields may overlap with the cavity in the neighboring nanobeam, producing optical coupling.
The optical coupling has a roughly decaying exponential behavior as a function of the separation between the nanobeams. 
We quantify the coupling strength between the two fundamental optical modes by looking at their odd and even combinations in the two nanobeam structures. 
Let us define the frequency of the odd (even) mode as $\lambda_{+}$ ($\lambda_{-}$). 
Its value is above (below) the frequency of a single beam fundamental mode $\lambda_{0}$ and we approximate it as $\lambda_{+} = \lambda_{0} + \gamma$ ($\lambda_{-} = \lambda_{0} - \gamma$), where the parameter $\gamma$ is the optical coupling strength. 
Our finite element model provides us with numerical values for the even and odd frequencies at various nanobeam separation values $s$ and, in consequence, allows us to extract an optical coupling strength $\gamma(s)$ as a function of the separation, Fig \ref{fig:2}. 
As expected, we find an exponential decay of the optical coupling strength as the separation between the beams $s$ increases. 
For the on-top configuration, we find simple exponential decay, Fig \ref{fig:2}(a) , for the optical coupling as a function of $s$ and, in contrast, the side-by-side configuration follows fourth-order exponential decay in $s$, Fig \ref{fig:2}(b). 
Additionally, the on-top configuration provides stronger optical coupling than the side-by-side configuration but might be experimentally difficult to fabricate.
The latter provides a much weaker optical coupling but its fabrication is more feasible \cite{Deotare2009}.

\section{Quantum mechanical model} \label{sec:S3}
The quantum mechanical description for our optomechanical system, composed by two isolated mechanical resonators, each supporting an optical mode with evanescent coupling between them,
\begin{align}
	\frac{\hat{H}}{\hbar} =  \sum_{j=1}^{2} \left[ \omega_{j} \hat{a}_{j}^{\dagger} \hat{a}_{j} + \nu_{j} \hat{b}_{j}^{\dagger} \hat{b}_{j} - g_{j} \hat{a}_{j}^{\dagger} \hat{a}_{j} \left( \hat{b}_{j}^{\dagger} + \hat{b}_{j}  \right) +  \Omega_{j} \cos \left( \omega_{dj} t\right) \left( \hat{a}_{j}^{\dagger} + \hat{a}_{j}  \right) \right] + \gamma \left( \hat{a}_{1}^{\dagger} \hat{a}_{2} + \hat{a}_{1} \hat{a}_{2}^{\dagger} \right),
 \label{opticalmechcoupling}
\end{align}
is given in terms of the creation (annihilation) operators for the optical $\hat{a}_{j}^{\dagger}$ ($\hat{a}_{j}$) and mechanical $\hat{b}_{j}^{\dagger}$ ($\hat{b}_{j}$) modes, the optical and mechanical mode frequencies are $\omega_{j}$ and $\nu_{j}$, in that order, the optomechanical coupling, optical driving, and optical coupling strengths are $g_{j}$, $\Omega_{j}$, and $\gamma$, respectively, the driving fields frequencies are $\omega_{d_{j}}$ with $j=1,2$. 
Moving into the reference frame defined by free optical fields,  $\vert\psi_{0}\rangle=e^{-i\sum_{j}\omega_{j}\hat{a}_{j}^{\dagger}\hat{a}_{j}t}\vert\psi_{1}\rangle$, allows us to apply a rotating wave approximation to disregard terms moving at fast optical frequencies, $\omega_{j} + \omega_{d_{j}}$, and consider an effective Hamiltonian,
\begin{align}
	\frac{\hat{H}_{1}}{\hbar} \approx  \sum_{j=1}^{2} \left[ \nu_{j} \hat{b}_{j}^{\dagger} \hat{b}_{j} - g_{j} \hat{a}_{j}^{\dagger} \hat{a}_{j} \left( \hat{b}_{j}^{\dagger} + \hat{b}_{j}  \right) +  \frac{1}{2}\Omega_{j} \left( \hat{a}_{j}^{\dagger}  e^{i \Delta_{j} t}  + \hat{a}_{j}   e^{-i \Delta_{j} t}  \right) \right] + \gamma \left[ \hat{a}_{1}^{\dagger} \hat{a}_{2} e^{i \delta t} + \hat{a}_{1} \hat{a}_{2}^{\dagger}e^{-i \delta t} \right],
\end{align}
where the optical driving term depends on the detuning between the resonant and driving frequencies, $\Delta_{j} = \omega_{j} - \omega_{dj}$, and the coupling terms by the detuning between resonant frequencies, $\delta = \omega_{1}-\omega_{2} $.
Now, a displacement of the mechanical modes proportional to the excitation number in the optical modes followed by moving to the frame defined by the free mechanical term, $\vert\psi_{1}\rangle =  e^{-\sum_{j} \alpha_{j} \hat{a}_{j}^{\dag}\hat{a}_{j} \left( \hat{b}_{j}^{\dag} - \hat{b}_{j} \right)}  e^{-i\sum_{j} \nu_{j}\hat{b}_{j}^{\dag}\hat{b}_{j} t} \vert\psi_{2}\rangle$ with $\alpha_{j} = -g_{j} / \nu_{j}$, yields an effective Hamiltonian,
\begin{align}
	\frac{\hat{H}_{2}}{\hbar}=\hat{H}_{K}+\hat{H}_{om}+\hat{H}_{oc}
\end{align}
with three components, an effective Kerr term, 
\begin{align}
	\hat{H}_{K}=-\sum_{j}\frac{g_{j}^{2}}{\nu_{j}}\left(\hat{a}_{j}^{\dagger}\hat{a}_{j}\right)^{2},
\end{align}
for the optical modes dependent on the ratio between the optomechanical coupling squared to the mechanical resonant frequency.
The optomechanical detuning term converts into parametric coupling between each mechanical resonator mode and its corresponding inscribed optical cavity mode,
\begin{align}
	\hat{H}_{OM} =& \frac{1}{2} \sum_{j=1}^{2} \Omega_{j} e^{-\frac{1}{2} \alpha_{j}^{2}} \sum_{p,q=0}^{\infty}\frac{(-1)^{q} \alpha_{j}^{p+q}}{p!q!}\left[ \hat{a}_{j}^{\dagger}\hat{b}_{j}^{\dagger p}\hat{b}_{j}^{q}e^{i\left[\Delta_{j}+\left(p-q\right)\nu_{j}\right]t} + \hat{a}_{j}\hat{b}_{j}^{\dagger q}\hat{b}_{j}^{p}e^{-i\left[\Delta_{j}+\left(p-q\right)\nu_{j}\right]t} \right] ,
\end{align}
feasible of control by the detuning between the external driving and the optical cavity. 
The optical coupling term converts into parametric coupling between optical and mechanical modes, 
\begin{align}
	\hat{H}_{OC} =& \gamma e^{-\frac{1}{2}\left( \alpha_{1}^{2} + \alpha_{2}^{2} \right)} \sum_{r,s,u,v=0}^{\infty}\frac{\left(-1\right)^{s+u}\alpha_{1}^{r+s}\alpha_{2}^{u+v}}{r!s!u!v!}\left[\hat{a}_{1}^{\dagger}\hat{a}_{2}\hat{b}_{1}^{\dagger r}\hat{b}_{1}^{s}\hat{b}_{2}^{\dagger u}\hat{b}_{2}^{v}e^{i\left[\delta+\left(r-s\right)\nu_{1}+\left(u-v\right)\nu_{2}\right]t}\right.\nonumber\\
 &\left.+\hat{a}_{1}\hat{a}_{2}^{\dagger}\hat{b}_{1}^{\dagger s}\hat{b}_{1}^{r}\hat{b}_{2}^{\dagger v}\hat{b}_{2}^{u}e^{-i\left[\delta+\left(r-s\right)\nu_{1}+\left(u-v\right)\nu_{2}\right]t}\right],
\end{align}
feasible of control by the detuning between the optical cavities resonant frequencies, $\delta = \omega_{1} - \omega_{2}$. 
Thus, we may control the parametric processes between each mechanical resonator and its inscribed optical mode via external driving fields, aiming for $\Delta_{j} + (p-q) \nu_{j} = 0$, but the parametric processes between isolated mechanical modes, mediated by the coupled optical modes, is controlled by the detuning between the optical cavities resonant frequencies, aiming for $\delta + (r-s) \nu_{1} + (u-v) \nu_{2}  =0$, which is provided by the fabrication itself.

\section{Mechanically controlled optical beam splitter} \label{sec:S4}
Let us drive the optical cavities on-resonance, $\Delta_{j} = 0$, such that the optomechanical coupling terms satisfy the optical pumping control condition $\Delta + (p-q) \nu_{j} = 0$ with $p=q$. 
In addition, if we consider the optical cavities identical, $\delta = 0$, such that the optical coupling terms with $r=s$ and $u=v$ satisfy the condition $\delta + (r-s) \nu_{1} + (u-v) \nu_{2}  =0$, we end up with a driven nonlinear optical beam splitter Hamiltonian,
\begin{align}\label{eq:opticalbs}
    \hat{H}_{NBS} = \sum_{j}\left\{-\frac{g_{j}^{2}}{\nu_{j}}\left(\hat{a}_{j}^{\dagger}\hat{a}_{j}\right)^{2} + \frac{\Omega_{j}}{2} \hat{F}\left[j,1,0\right] \left(  \hat{a}_{j}^{\dagger} +  \hat{a}_{j}  \right)\right\} +\gamma \hat{F}\left[1,1,0\right] \hat{F}\left[2,1,0\right]
    \left( \hat{a}_{1}^{\dagger} \hat{a}_{2}  +  \hat{a}_{1} \hat{a}_{2}^{\dagger}  \right),
\end{align}
where the driving strength and the optical coupling strength depend on the auxiliary Hermitian operator function,
\begin{align}
    \hat{F}\left[j,p,q\right]= \left( -\frac{g_{j}}{\nu_{j}} \right)^{q} e^{-\frac{1}{2}\left( \frac{g_{j}}{\nu_{j}} \right)^{2}} \mathrm{_1F}_1\left[-\hat{b}_{j}^{\dagger}\hat{b}_{j}; p; \left( \frac{g_{j}}{\nu_{j}} \right)^{2}\right],
 \end{align}
 given in terms of the optomechanical coupling strength $g_{j}$, the resonant mechanical frequency $\nu_{j}$, the mechanical excitation number $\hat{b}_{j}^{\dagger} \hat{b}_{j}$, and the confluent hypergeometric function $ \mathrm{_1F}_1\left[a; b; z\right]$. 
 
 For the typical optomechanical coupling strength to resonant frequency ratio in nanobeams, $g_{j}/\nu_{j}\ll1$, the auxiliary Hermitian operator function may be approximated to a form,
 \begin{equation}\label{eq:auxiliaryoperatorf}
 \hat{F}\left[j,p,q\right]\simeq\left(\frac{g_{j}}{\nu_{j}} \right)^{q}  \left[1-\left(\frac{1}{2}+\frac{\hat{b}_{j}^{\dagger}\hat{b}_{j}}{p}\right)\left( \frac{g_{j}}{\nu_{j}} \right)^{2}+\mathcal{O}\left( \frac{g_{j}}{\nu_{j}} \right)^{4}\right],
 \end{equation}
whose leading order depends on the mechanical excitation number $\hat{b
}_{j}^{\dagger} \hat{b}_{j}$; for example, 
 \begin{equation}
 \hat{F}\left[j,1,0\right] \approx 1 - \left(\frac{1}{2} + \hat{b}_{j}^{\dagger}\hat{b}_{j} \right) \left( \frac{g_{j}}{\nu_{j}} \right)^{2},
 \end{equation}
 a sufficiently small mechanical excitation number, $ \hat{F}\left[j,1,0\right] \approx 1$, provides us with an effective driven nonlinear optical beam splitter,
 \begin{align}
    \hat{H}_{NBS} \approx \sum_{j}\left\{-\frac{g_{j}^{2}}{\nu_{j}}\left(\hat{a}_{j}^{\dagger}\hat{a}_{j}\right)^{2} + \frac{\Omega_{j}}{2}  \left(  \hat{a}_{j}^{\dagger} +  \hat{a}_{j}  \right) \right\}+ \gamma  
    \left( \hat{a}_{1}^{\dagger} \hat{a}_{2}  +  \hat{a}_{1} \hat{a}_{2}^{\dagger}  \right),
\end{align}
with constant parameters in the so-called Rabi regime, $g_{j}^{2} / \nu_{j} \ll \gamma$, where the spectrum of the system without driving, $\Omega=0$, is linear.
As a result, we may approximate the system,
\begin{align}
    \hat{H}_{NBS} \approx  \sum_{j}\frac{\Omega_{j}}{2}  \left(  \hat{a}_{j}^{\dagger} +  \hat{a}_{j}  \right) + \gamma  
    \left( \hat{a}_{1}^{\dagger} \hat{a}_{2}  +  \hat{a}_{1} \hat{a}_{2}^{\dagger}  \right),
\end{align}
and recover a standard optical beam splitter with driving.
In the future, it may be possible to have larger optomechanical coupling strength that allows exploring the nonlinear regimes available in the model. 
Our mechanically isolated configuration may explore these nonlinear regimes in the case where the nanobeams are sufficiently apart from each other, such that the optical coupling is minimal, making it a trivial scenario. 

\begin{figure}
\includegraphics{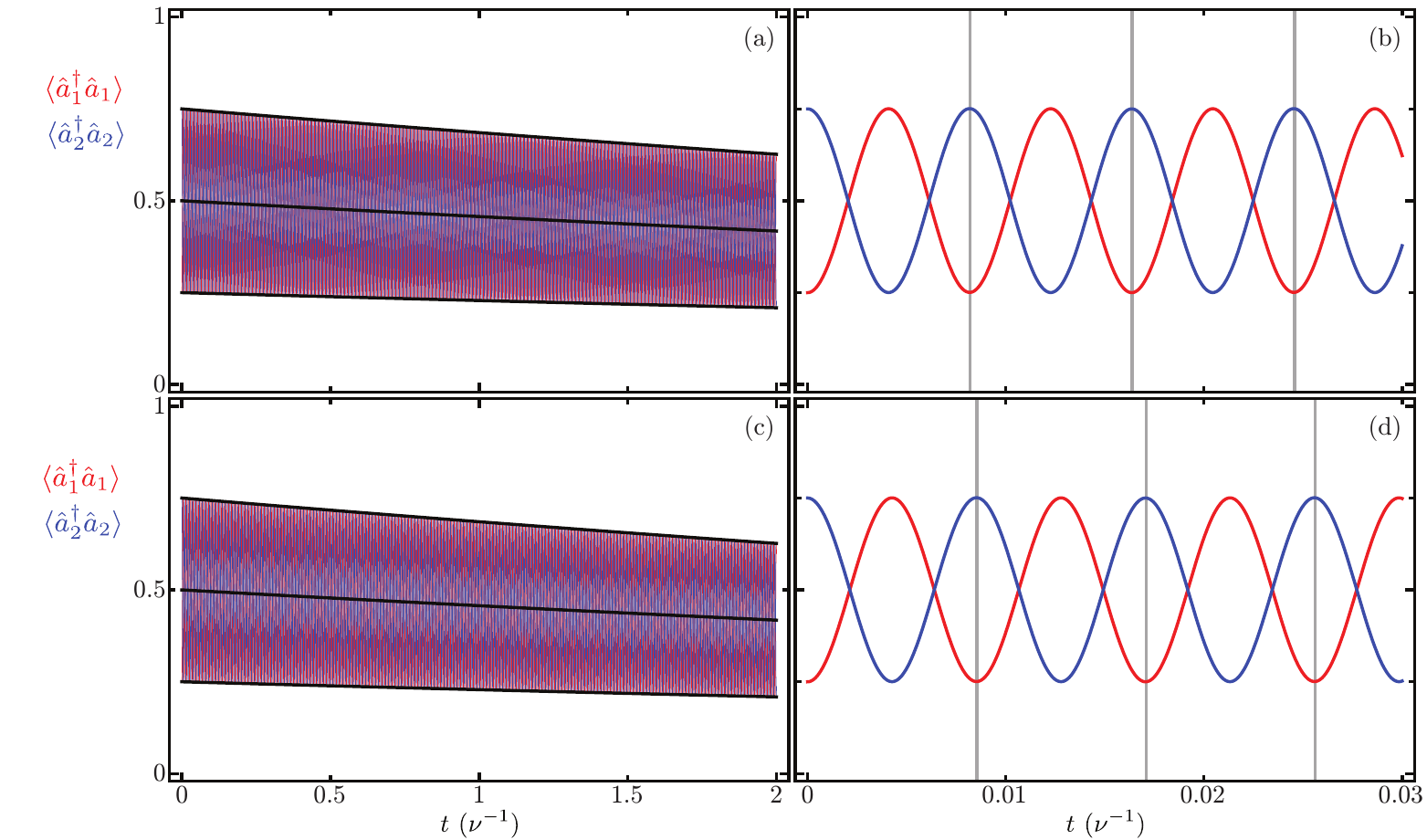}
\caption{Time evolution of the optical mode excitation number, $ \langle \hat{a}_{j}^{\dagger} \hat{a}_{j} \rangle$ with $j = 1, 2$, for initial states  (a)--(b)
$\left\vert \psi (0) \right\rangle = \left[ \cos (\frac{\pi}{3}) \left\vert 1,0 \right\rangle_{\mathrm{opt}} + \sin \left(\frac{\pi}{3}\right) \left\vert 0,1 \right\rangle_{\mathrm{opt}} \right] \left\vert 0,0 \right\rangle_{\mathrm{mec}}$ and (c)--(d) $\left\vert \psi (0) \right\rangle = \left[ \cos \left(\frac{\pi}{3}\right) \left\vert 1,0 \right\rangle_{\mathrm{opt}} + \sin \left(\frac{\pi}{3}\right) \left\vert 0,1 \right\rangle_{\mathrm{opt}} \right] \left\vert 0,1 \right\rangle_{\mathrm{mec}}$ for (a) and (c) long 
 and (b) and (c) short evolution times. The exchange period between optical modes is indicated by vertical gray lines.}\label{fig:3}
\end{figure}

Figure \ref{fig:3} shows the evolution of the mean value of the optical mode excitation number, $ \langle \hat{a}_{j}^{\dagger} \hat{a}_{j} \rangle$ with $j = 1, 2$ using the full master equation in the simplified reference frame provided by our Hamiltonian in Eq.~(\ref{eq:opticalbs}) under the leading order approximation for the auxiliary Hermitian operator function $\hat{F}[j,p,q]$ in Eq.~(\ref{eq:auxiliaryoperatorf}). 
We must emphasize that optical excitation numbers remain unchanged by the reference frame transformations.
While mechanical excitation numbers are affected by these reference frame transformations, the numerical difference between the simplified and laboratory reference frames remains numerically small of the order of $\lesssim 10^{-7}$.
In the numerical simulation, we introduce optical and mechanical losses given by  $\gamma_{o,j} = 0.09/\nu$ and $\gamma_{m,j} = 1.5 \times 10^{-5} / \nu$, consistent with experimental results \cite{Chan2011}. For the optical component of the initial state, we use initial states with one excitation entangled between the two optical modes, to produce oscillations between them. 
For the mechanical component of the initial state, we use two distinct states to compare the effect of mechanical excitation numbers. 
One initial state has zero mechanical excitation 
$\left\vert \psi (0) \right\rangle = \left[ \cos \left(\frac{\pi}{3}\right) \left\vert 1,0 \right\rangle_{\mathrm{opt}} + \sin \left(\frac{\pi}{3}\right) \left\vert 0,1 \right\rangle_{\mathrm{opt}} \right] \left\vert 0,0 \right\rangle_{\mathrm{mec}}$, Fig. \ref{fig:3}(a) and Fig. \ref{fig:3}(b). 
The other initial state has one mechanical excitation
$\left\vert \psi (0) \right\rangle = \left[ \cos \left(\frac{\pi}{3}\right) \left\vert 1,0 \right\rangle_{\mathrm{opt}} + \sin \left(\frac{\pi}{3}\right) \left\vert 0,1 \right\rangle_{\mathrm{opt}} \right] \left\vert 0,1 \right\rangle_{\mathrm{mec}}$, Fig. \ref{fig:3}(c) and Fig. \ref{fig:3}(d).
For long evolution times, we observe decay due to optical losses, Fig. \ref{fig:3}(a) and Fig. \ref{fig:3}(c). 
For short evolution times, Fig. \ref{fig:3}(b) and Fig. \ref{fig:3}(d), we observe optical excitation exchange with temporal period, 
\begin{align}
\tau = 2 \pi \left\{ \gamma  e^{-\frac{1}{2}\left( \frac{g_{1}}{\nu_{1}} \right)^{2}} e^{-\frac{1}{2}\left( \frac{g_{2}}{\nu_{2}} \right)^{2}} \mathrm{_1F}_1\left[ - \langle \hat{b}_{1}^{\dagger} \hat{b}_{1} \rangle; 1; \left( \frac{g_{1}}{\nu_{1}} \right)^{2}\right] \mathrm{_1F}_1\left[- \langle \hat{b}_{2}^{\dagger} \hat{b}_{2} \rangle ; 1; \left( \frac{g_{2}}{\nu_{2}} \right)^{2}\right] \right\}^{-1},
\end{align} 
that depends on the mechanical excitation number, $\langle \hat{b}_{j}^{\dagger} \hat{b}_{j} \rangle $ with $j=1,2$.

\section{Optically controlled mechanical coupler} \label{sec:S5}
Let us drive the red sideband of the optical cavities, $\Delta_{j} = \nu_{j}$, such that the optomechanical coupling terms with $q=p+1$ satisfy the optical pumping control condition $\Delta_{j} + (p-q) \nu_{j} = 0$. 
In addition, we consider the optical cavities with a detunning equivalent to $\delta = -\nu_{1} + \nu_{2} $, such that the optical coupling terms with $r=s+1$ and $v=u+1$ satisfy the condition $\delta + (r-s) \nu_{1} + (u-v) \nu_{2}  =0$.
Under these considerations, the effective Hamiltonian describing the system,
\begin{align}\label{eq:nonlinearcoupler}
    \hat{H}_{OMC} =& -\sum_{j}\left\{\frac{g_{j}^{2}}{\nu_{j}}\left(\hat{a}_{j}^{\dagger}\hat{a}_{j}\right)^{2} + \frac{\Omega_{j}}{2}\left( \hat{a}_{j}^{\dagger}F\left[j,2,1\right] \hat{b}_{j} +  \hat{a}_{j} \hat{b}_{j}^{\dagger}F\left[j,2,1\right] \right)\right\} \nonumber 
 \\
    &  -\gamma\left\{\hat{a}_{1}^{\dagger} \hat{a}_{2} \hat{b}_{1}^{\dagger} \hat{F}\left[1,2,1\right] \hat{F}\left[2,2,1\right] \hat{b}_{2}  +  \hat{a}_{1} \hat{a}_{2}^{\dagger}\hat{b}_{2}^{\dagger}  \hat{F}\left[1,2,1\right] \hat{F}\left[2,2,1\right]  \hat{b}_{1}\right\}
\end{align}
becomes a nonlinear coupler of mechanical and optical modes where the excitation transfer between optical modes is accompanied by the transfer of mechanical excitation. 
Here, we used the auxiliary Hermitian operator function $\hat{F}[j, p, q]$ defined before.

A first-order approximation using the coupling and pump rates ranges available in the experimental setups discussed in Sec. \ref{sec:S2} leads to the following effective model,
\begin{align}
    \hat{H}_{OMC} \approx& -\sum_{j}\left[\frac{g_{j}^{2}}{\nu_{j}}\left(\hat{a}_{j}^{\dagger}\hat{a}_{j}\right)^{2} + \Omega_{eff_{j}} \left(  \hat{a}_{j}^{\dagger} \hat{b}_{j} +  \hat{a}_{j} \hat{b}_{j}^{\dagger} \right)\right] -  \Gamma_{eff} \left( \hat{a}_{1}^{\dagger} \hat{a}_{2} \hat{b}_{1}^{\dagger} \hat{b}_{2}  +  \hat{a}_{1} \hat{a}_{2}^{\dagger}  \hat{b}_{2}^{\dagger}  \hat{b}_{1}\right),
\end{align}
with the effective linear optomechanical coupling strength $\Omega_{eff_{j}}= \Omega g_{j}/ (2 \nu_{j})$ and the parametric coupling strength $\Gamma_{eff}=\gamma g_{1}g_{2}/(\nu_{1}\nu_{2})$ mixing all optical and mechanical modes. 
For the nanobeams under consideration and using the maximum feasible pump rate of $10^{5}~\mathrm{nW}$, $\Omega_{eff_{j}}$ becomes the leading coupling, which is of the order of tens of megahertz. 
The second leading rate is $\Gamma_{eff}$, which is of the order of a few megahertz or fractions of a megahertz. 
Finally, the coupling $g^{2}_{j}/\nu_{j}$ is the smallest of the three, of the order of a few kilohertz.
Such that we may approximate,
\begin{align}
    \hat{H}_{OMC} \approx& -\sum_{j}  \Omega_{eff_{j}} \left(  \hat{a}_{j}^{\dagger} \hat{b}_{j} +  \hat{a}_{j} \hat{b}_{j}^{\dagger} \right) -  \Gamma_{eff} \left( \hat{a}_{1}^{\dagger} \hat{a}_{2} \hat{b}_{1}^{\dagger} \hat{b}_{2}  +  \hat{a}_{1} \hat{a}_{2}^{\dagger}  \hat{b}_{2}^{\dagger}  \hat{b}_{1}\right),
\end{align}
the dynamics with a linear parametric model that associates the exchange of optical excitation with that of mechanical excitation.

\begin{figure}
\includegraphics{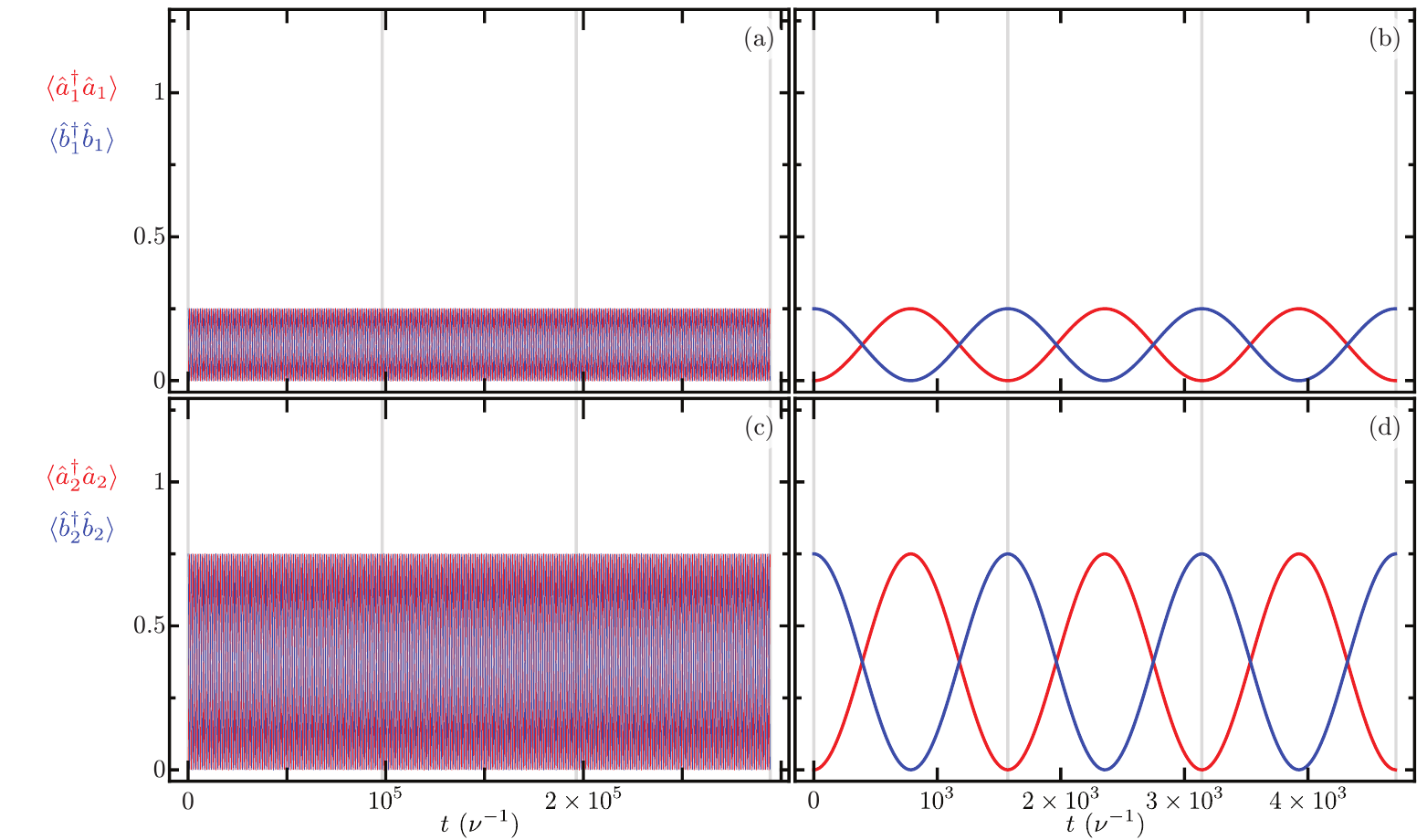}
\caption{Time evolution of the optical, $ \langle \hat{a}_{j}^{\dagger} \hat{a}_{j} \rangle$, and mechanical, $ \langle \hat{b}_{j}^{\dagger} \hat{b}_{j} \rangle$, modes excitation number at each nanobeam, (a)--(b) $j=1$ and (b)--(c) $j=2$, for  (a) and (c) long,  and (b) and (d) short evolution times. The initial state is $\left\vert \psi (0) \right\rangle =  \left\vert 0,0 \right\rangle_{\mathrm{opt}} \left[ \cos \left(\frac{\pi}{3}\right) \left\vert 1,0 \right\rangle_{\mathrm{opt}} + \sin \left(\frac{\pi}{3}\right) \left\vert 0,1 \right\rangle_{\mathrm{mec}} \right]$. The exchange period between (a) and (c) optomechanical modes in both nanobeams and (b) and (d) optomechanical modes in each nanobean is indicated by vertical gray lines.}\label{fig:4}
\end{figure}

\begin{figure}
\includegraphics{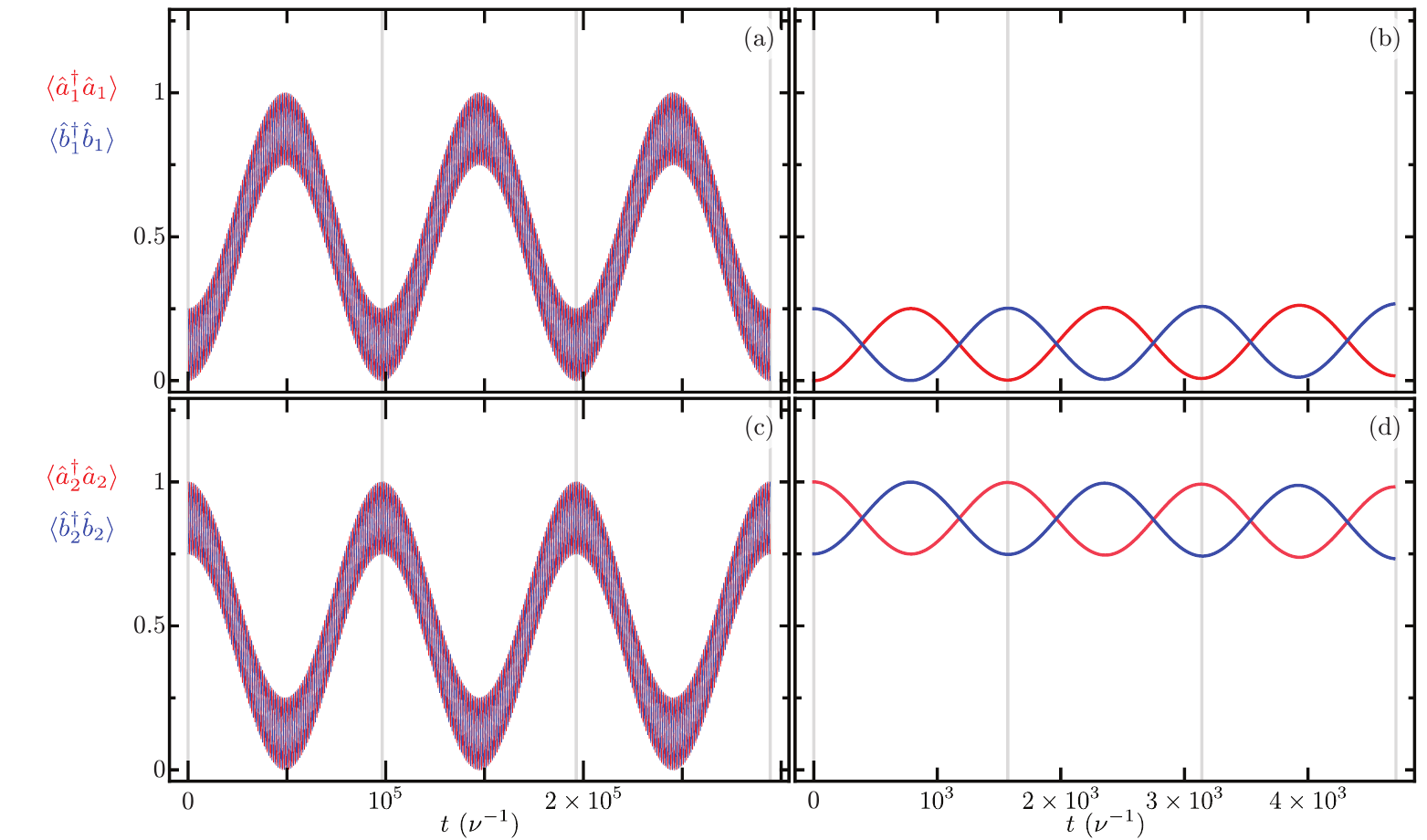}
\caption{Same as Fig. \ref{fig:4} with initial state $\left\vert \psi (0) \right\rangle =  \left\vert 0,1 \right\rangle_{\mathrm{opt}} \left[ \cos \left(\frac{\pi}{3}\right) \left\vert 1,0 \right\rangle_{\mathrm{opt}} + \sin \left(\frac{\pi}{3}\right) \left\vert 0,1 \right\rangle_{\mathrm{mec}} \right]$.}\label{fig:5}
\end{figure}

Figure \ref{fig:4} (Figure \ref{fig:5}) show the Schr\"odinger equation time evolution of the optomechanical excitation numbers in both nanobeams under the effective nonlinear coupled Hamiltonian in Eq.~(\ref{eq:nonlinearcoupler}). Where, again, we use the leading order approximation for the auxiliary Hermitian operator function $\hat{F}[j,p,q]$ in Eq.~(\ref{eq:auxiliaryoperatorf}). As in the previous section, the numerical effect of frame changes on the mean value of excitation numbers is negligible.
For short, Fig \ref{fig:4}(a) and Fig \ref{fig:4}(c) (Fig \ref{fig:5}(a) and Fig \ref{fig:5}(c)),   and long, Fig \ref{fig:4}(b) and Fig \ref{fig:4}(d) (Fig \ref{fig:5}(b) and Fig \ref{fig:5}(d)), evolution times with an initial state with no excitation in the optical modes, $\left\vert \psi (0) \right\rangle =  \left\vert 0,0 \right\rangle_{\mathrm{opt}} \left[ \cos \left(\frac{\pi}{3}\right) \left\vert 1,0 \right\rangle_{\mathrm{opt}} + \sin \left(\frac{\pi}{3}\right) \left\vert 0,1 \right\rangle_{\mathrm{mec}} \right]$ $ \left( \left\vert \psi (0) \right\rangle =  \left\vert 0,1 \right\rangle_{\mathrm{opt}} \left[ \cos \left(\frac{\pi}{3}\right) \left\vert 1,0 \right\rangle_{\mathrm{opt}} + \sin \left(\frac{\pi}{3}\right) \left\vert 0,1 \right\rangle_{\mathrm{mec}} \right] \right)$.
Figure \ref{fig:4}(b) and Figure \ref{fig:4}(d) (Figure \ref{fig:5}(b) and Figure \ref{fig:5}(d)) show the predicted excitation exchange between the optical and mechanical modes in each nanobeam with frequency exchange temporal period, 
\begin{align}
    \tau_{\mathrm{om}, j} = 2 \pi \left\{ \frac{\Omega_{j}}{2}  \left( -\frac{g_{j}}{\nu_{j}} \right) e^{-\frac{1}{2}\left( \frac{g_{j}}{\nu_{j}} \right)^{2}} \mathrm{_1F}_1\left[- \langle \hat{b}_{j}^{\dagger}\hat{b}_{j} \rangle; 2; \left( \frac{g_{j}}{\nu_{j}} \right)^{2}\right] \right\}^{-1},
\end{align} \label{eq:omperiod}
that induces the exchange of mechanical excitation with a period,
\begin{align}
    \tau_{ \mathrm{mec} } = 2 \pi \left\{ \gamma  \left( -\frac{g_{1} g_{2}}{\nu_{1} \nu_{2}} \right) e^{-\frac{1}{2}\left( \frac{g_{1}}{\nu_{1}} \right)^{2}} e^{-\frac{1}{2}\left( \frac{g_{2}}{\nu_{2}} \right)^{2}} \mathrm{_1F}_1\left[-\langle \hat{b}_{1}^{\dagger}\hat{b}_{1} \rangle; 2; \left( \frac{g_{1}}{\nu_{1}}  \right)^{2}\right] \mathrm{_1F}_1\left[-\langle \hat{b}_{2}^{\dagger}\hat{b}_{2} \rangle; 2; \left( \frac{g_{2}}{\nu_{2}}  \right)^{2}\right] \right\}^{-1},
\end{align}
that can be observed in Fig. \ref{fig:4}(a) and Fig. \ref{fig:4}(c). (Fig. \ref{fig:5}(a) and Fig. \ref{fig:5}(c)). 
For the sake of illustration, this simulation was performed with a closed system with no losses.

\section{Optically controlled mechanical two-mode squeezing} \label{sec:S6}
Finally, let us drive the red sideband of the optical cavities, $\Delta_{j} = \nu_{j}$, such that the optomechanical coupling terms with $q=p+1$ satisfy the optical pumping control condition $\Delta + (p-q) \nu_{j} = 0$. 
In addition, we consider the optical cavities with a detunning equivalent to , $\delta = -\nu_{1} - \nu_{2} $, such that the optical coupling terms with $r=s+1$ and $u=v+1$ satisfy the condition $\delta + (r-s) \nu_{1} + (u-v) \nu_{2}  =0$.
Under these considerations, the effective Hamiltonian describing the system,
\begin{align}\label{eq:kerrplus}
    \hat{H}_{OMC} 
    =& -\sum_{j}\left\{\frac{g_{j}^{2}}{\nu_{j}}\left(\hat{a}_{j}^{\dagger}\hat{a}_{j}\right)^{2} + \frac{\Omega_{j}}{2}\left( \hat{a}_{j}^{\dagger}F\left[j,2,1\right] \hat{b}_{j} +  \hat{a}_{j} \hat{b}_{j}^{\dagger}F\left[j,2,1\right] \right)\right\} \nonumber 
 \\
    &  -\gamma\left\{\hat{a}_{1}^{\dagger} \hat{a}_{2} \hat{b}_{1}^{\dagger} \hat{b}_{2}^{\dagger}F\left[1,2,1\right] F\left[2,2,1\right]   +  \hat{a}_{1} \hat{a}_{2}^{\dagger}  F\left[1,2,1\right] F\left[2,2,1\right]  \hat{b}_{2}\hat{b}_{1}\right\}
\end{align}
becomes a more complex model where the first term is the standard nonlinear Kerr term, the second term is, again, linear optomechanical coupling at each nanobeam, and the third term suggest two-mode parametric coupling of the mechanical modes mediated by excitation exchange of the optical modes.
Again, we used the auxiliary Hermitian operator function $\hat{F}[j, p, q]$ defined before.

Again, a first-order approximation using the coupling and pump rates ranges available for current experimental setups leads to the following effective model,
\begin{align}
    \hat{H}_{OMS} \approx& -\sum_{j}\left[\frac{g_{j}^{2}}{\nu_{j}}\left(\hat{a}_{j}^{\dagger}\hat{a}_{j}\right)^{2} + \Omega_{eff_{j}} \left(  \hat{a}_{j}^{\dagger} \hat{b}_{j} +  \hat{a}_{j} \hat{b}_{j}^{\dagger} \right) \right]-  \Gamma_{eff} \left( \hat{a}_{1}^{\dagger} \hat{a}_{2} \hat{b}_{1}^{\dagger} \hat{b}_{2}^{\dagger} +  \hat{a}_{1} \hat{a}_{2}^{\dagger}    \hat{b}_{1}\hat{b}_{2} \right),
\end{align}
where the effective linear optomechanical coupling $\Omega_{eff_{j}}= \Omega g_{j}/ (2 \nu_{j})$ and the parametric coupling strength $\Gamma_{eff}=\gamma g_{1}g_{2}/(\nu_{1}\nu_{2})$ are equal to those defined before and follow an identical hierarchy that yields the approximate effective Hamiltonian, 
\begin{align}
    \hat{H}_{OMS} \approx& -\sum_{j} \Omega_{eff_{j}} \left(  \hat{a}_{j}^{\dagger} \hat{b}_{j} +  \hat{a}_{j} \hat{b}_{j}^{\dagger} \right) -  \Gamma_{eff} \left( \hat{a}_{1}^{\dagger} \hat{a}_{2} \hat{b}_{1}^{\dagger} \hat{b}_{2}^{\dagger} +  \hat{a}_{1} \hat{a}_{2}^{\dagger}    \hat{b}_{1}\hat{b}_{2} \right),
\end{align}
whose dynamics are that of two linearly coupled optomechanical systems with an extra term that associates the exchange of optical excitation with two-mode mechanical squeezing.

\begin{figure}
\includegraphics{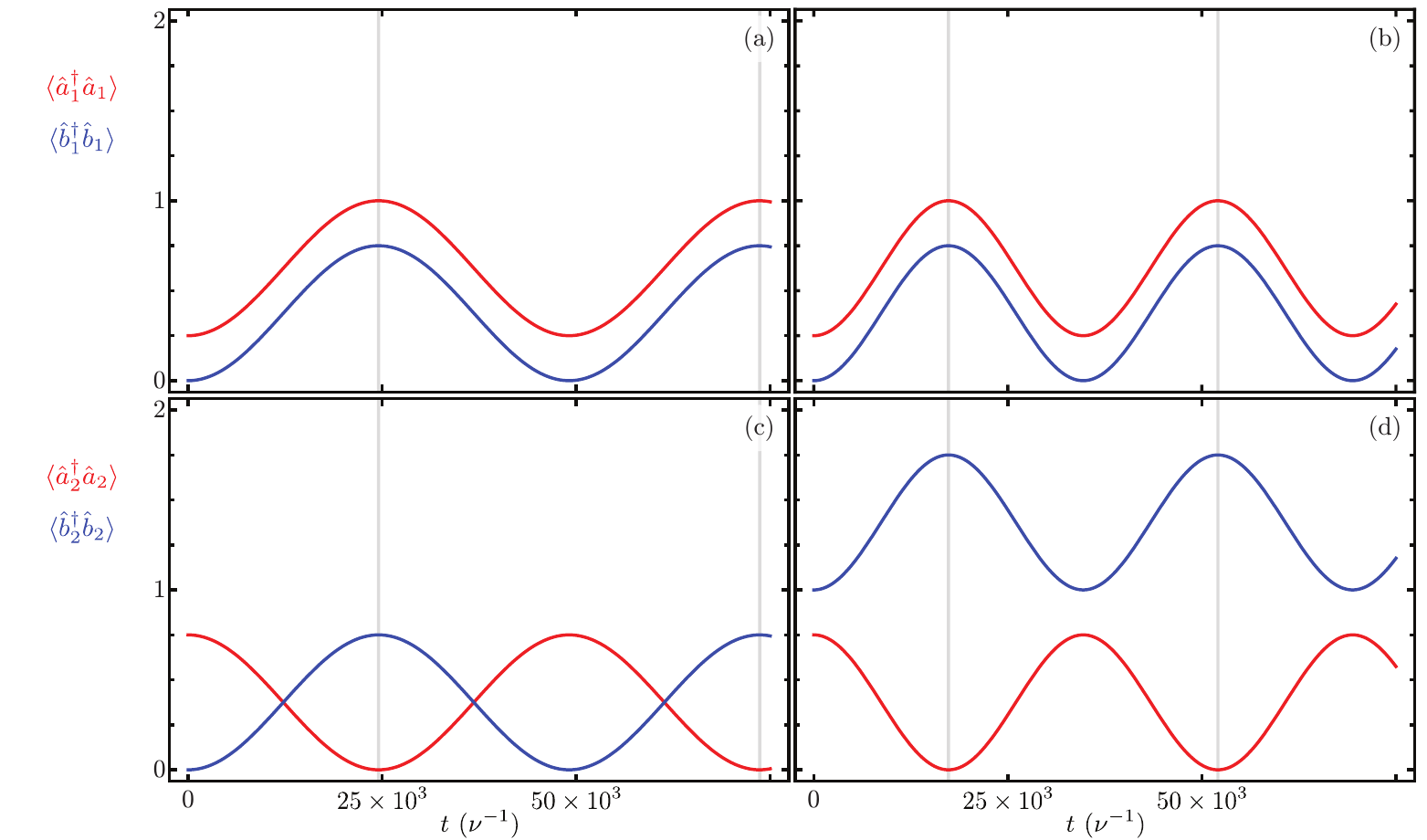}
\caption{Time evolution of optical, $ \langle \hat{a}_{j}^{\dagger} \hat{a}_{j} \rangle$, and mechanical, $ \langle \hat{b}_{j}^{\dagger} \hat{b}_{j} \rangle$, modes excitation number in each nanobeam, (a)--(b) $j = 1$ and (c)--(d) $j = 2$, for initial states (a) and (c)  
$\left\vert \psi (0) \right\rangle = \left[ \cos (\frac{\pi}{3}) \left\vert 1,0 \right\rangle_{\mathrm{opt}} + \sin \left(\frac{\pi}{3}\right) \left\vert 0,1 \right\rangle_{\mathrm{opt}} \right] \left\vert 0,0 \right\rangle_{\mathrm{mec}}$ and (b) and (d) $\left\vert \psi (0) \right\rangle = \left[ \cos \left(\frac{\pi}{3}\right) \left\vert 1,0 \right\rangle_{\mathrm{opt}} + \sin \left(\frac{\pi}{3}\right) \left\vert 0,1 \right\rangle_{\mathrm{opt}} \right] \left\vert 0,1 \right\rangle_{\mathrm{mec}}$ for short evolution times. The exchange period between optomechanical modes in each nanobeam is indicated by vertical gray lines.}\label{fig:6}
\end{figure}

Figure \ref{fig:6} shows the Schr\"odinger time evolution under the effective Hamiltonian in Eq. (\ref{eq:kerrplus}) and with the leading order approximation for the auxiliary Hermitian operator function $\hat{F}[j,p,q]$. These results, just like in the two previous sections display negligible differences in the mean excitation numbers due to frame changes.
 Figure \ref{fig:6}(a) and Fig. \ref{fig:6}(c) show short evolution times with an initial state with no excitation in the mechanical modes, $\left\vert \psi (0) \right\rangle =  \left\vert 0,0 \right\rangle_{\mathrm{opt}} \left[ \cos \left(\frac{\pi}{3}\right) \left\vert 1,0 \right\rangle_{\mathrm{opt}} + \sin \left(\frac{\pi}{3}\right) \left\vert 0,1 \right\rangle_{\mathrm{mec}} \right]$ and Fig. \ref{fig:6}(b) and Fig. \ref{fig:6}(d) for an initial state with a single excitation in the mechanical modes,  $\left\vert \psi (0) \right\rangle =  \left\vert 0,1 \right\rangle_{\mathrm{opt}} \left[ \cos \left(\frac{\pi}{3}\right) \left\vert 1,0 \right\rangle_{\mathrm{opt}} + \sin \left(\frac{\pi}{3}\right) \left\vert 0,1 \right\rangle_{\mathrm{mec}} \right]$, 
All figures show the predicted excitation exchange between optical and mechanical modes in each nanobeam with temporal period given by Eq.(\ref{eq:omperiod}). 
The period related to two-mode squeezing occurs at long evolution times that are not suitable of numerical simulation due to the increase in excitation number provided by the two-mode squeezing.
For the sake of illustration, this simulation was performed with a closed system with no losses.

\section{Conclusion} \label{sec:S7}
We proposed a Hamiltonian model composed of two mechanical vibrational modes and two optical modes. 
The vibrational modes are mechanically isolated and coupled to their corresponding optical mode under standard optomechanical interaction.
We allow for independent external driving and evanescent coupling of the optical modes.
We built a finite element model of the classical problem to recover the ranges of values for variables of interest; that is, mechanical and optical resonant frequencies for the isolated elements, their optical coupled modes, and the naked coupling strength corresponding to different configurations and separations.

We showed that our model allows the coupling of the isolated mechanical modes mediated by the optical fields. 
The difference between resonant frequencies of the optical modes, which may be hard to control in experimental setups, and between them and the external optical driving field frequencies control the type of mechanical interaction produced.
Thanks to the ranges of parameter values, we are able to approximate the models into linear models for a small number of excitation in the system that, for example, induces an optical beam splitter where the mechanical state dresses the optical coupling, a mechanical bidirectional coupler or a two-mode squeezer where the optical state of the system controls the interaction coupling strength.

\section*{Acknowledgments}	
F.~E.~O., B.~R.~J.-A. and F~.H~.M.-V. thank CONACYT for financial support. 
F.~E.~O. thanks U.~N.~N. for study leave support. 


\section*{References}

\begin{thebibliography}{40}%
\makeatletter
\providecommand \@ifxundefined [1]{%
 \@ifx{#1\undefined}
}%
\providecommand \@ifnum [1]{%
 \ifnum #1\expandafter \@firstoftwo
 \else \expandafter \@secondoftwo
 \fi
}%
\providecommand \@ifx [1]{%
 \ifx #1\expandafter \@firstoftwo
 \else \expandafter \@secondoftwo
 \fi
}%
\providecommand \natexlab [1]{#1}%
\providecommand \enquote  [1]{``#1''}%
\providecommand \bibnamefont  [1]{#1}%
\providecommand \bibfnamefont [1]{#1}%
\providecommand \citenamefont [1]{#1}%
\providecommand \href@noop [0]{\@secondoftwo}%
\providecommand \href [0]{\begingroup \@sanitize@url \@href}%
\providecommand \@href[1]{\@@startlink{#1}\@@href}%
\providecommand \@@href[1]{\endgroup#1\@@endlink}%
\providecommand \@sanitize@url [0]{\catcode `\\12\catcode `\$12\catcode
  `\&12\catcode `\#12\catcode `\^12\catcode `\_12\catcode `\%12\relax}%
\providecommand \@@startlink[1]{}%
\providecommand \@@endlink[0]{}%
\providecommand \url  [0]{\begingroup\@sanitize@url \@url }%
\providecommand \@url [1]{\endgroup\@href {#1}{\urlprefix }}%
\providecommand \urlprefix  [0]{URL }%
\providecommand \Eprint [0]{\href }%
\providecommand \doibase [0]{https://doi.org/}%
\providecommand \selectlanguage [0]{\@gobble}%
\providecommand \bibinfo  [0]{\@secondoftwo}%
\providecommand \bibfield  [0]{\@secondoftwo}%
\providecommand \translation [1]{[#1]}%
\providecommand \BibitemOpen [0]{}%
\providecommand \bibitemStop [0]{}%
\providecommand \bibitemNoStop [0]{.\EOS\space}%
\providecommand \EOS [0]{\spacefactor3000\relax}%
\providecommand \BibitemShut  [1]{\csname bibitem#1\endcsname}%
\let\auto@bib@innerbib\@empty
\bibitem [{\citenamefont {Dorsel}\ \emph {et~al.}(1983)\citenamefont {Dorsel},
  \citenamefont {Mc{C}ullen}, \citenamefont {Meystre}, \citenamefont {Vignes},\
  and\ \citenamefont {Walther}}]{Dorsel1983}%
  \BibitemOpen
  \bibfield  {author} {\bibinfo {author} {\bibfnamefont {A.}~\bibnamefont
  {Dorsel}}, \bibinfo {author} {\bibfnamefont {J.~D.}\ \bibnamefont
  {Mc{C}ullen}}, \bibinfo {author} {\bibfnamefont {P.}~\bibnamefont {Meystre}},
  \bibinfo {author} {\bibfnamefont {E.}~\bibnamefont {Vignes}},\ and\ \bibinfo
  {author} {\bibfnamefont {H.}~\bibnamefont {Walther}},\ }\bibfield  {title}
  {\bibinfo {title} {Optical bistability and mirror confinement induced by
  radiation pressure},\ }\href {https://doi.org/10.1103/Phys.Rev.Lett.51.155}
  {\bibfield  {journal} {\bibinfo  {journal} {Phys. Rev. Lett.}\ }\textbf
  {\bibinfo {volume} {51}},\ \bibinfo {pages} {155} (\bibinfo {year}
  {1983})}\BibitemShut {NoStop}%
\bibitem [{\citenamefont {Aldana}\ \emph {et~al.}(2013)\citenamefont {Aldana},
  \citenamefont {Bruder},\ and\ \citenamefont {Nunnenkamp}}]{Aldana2013}%
  \BibitemOpen
  \bibfield  {author} {\bibinfo {author} {\bibfnamefont {S.}~\bibnamefont
  {Aldana}}, \bibinfo {author} {\bibfnamefont {C.}~\bibnamefont {Bruder}},\
  and\ \bibinfo {author} {\bibfnamefont {A.}~\bibnamefont {Nunnenkamp}},\
  }\bibfield  {title} {\bibinfo {title} {Equivalence between an optomechanical
  system and a {K}err medium},\ }\href
  {https://doi.org/10.1103/PhysRevA.88.043826} {\bibfield  {journal} {\bibinfo
  {journal} {Phys. Rev. A}\ }\textbf {\bibinfo {volume} {88}},\ \bibinfo
  {pages} {043826} (\bibinfo {year} {2013})},\ \Eprint
  {https://arxiv.org/abs/1306.0415} {arXiv:1306.0415 [quant-ph]} \BibitemShut
  {NoStop}%
\bibitem [{\citenamefont {Cuthbertson}\ \emph {et~al.}(1996)\citenamefont
  {Cuthbertson}, \citenamefont {Tobar}, \citenamefont {Ivanov},\ and\
  \citenamefont {Blair}}]{Cuthbertson1996}%
  \BibitemOpen
  \bibfield  {author} {\bibinfo {author} {\bibfnamefont {B.~D.}\ \bibnamefont
  {Cuthbertson}}, \bibinfo {author} {\bibfnamefont {M.~E.}\ \bibnamefont
  {Tobar}}, \bibinfo {author} {\bibfnamefont {E.~N.}\ \bibnamefont {Ivanov}},\
  and\ \bibinfo {author} {\bibfnamefont {D.~G.}\ \bibnamefont {Blair}},\
  }\bibfield  {title} {\bibinfo {title} {Parametric back-action effects in a
  high-{Q} cryogenic sapphire transducer},\ }\href
  {https://doi.org/https://doi.org/10.1063/1.1147193} {\bibfield  {journal}
  {\bibinfo  {journal} {Rev. Sci. Instrum.}\ }\textbf {\bibinfo {volume}
  {67}},\ \bibinfo {pages} {2435} (\bibinfo {year} {1996})}\BibitemShut
  {NoStop}%
\bibitem [{\citenamefont {Aspelmeyer}\ \emph {et~al.}(2014)\citenamefont
  {Aspelmeyer}, \citenamefont {Kippenberg},\ and\ \citenamefont
  {Marquardt}}]{Aspelmeyer2014}%
  \BibitemOpen
  \bibfield  {author} {\bibinfo {author} {\bibfnamefont {M.}~\bibnamefont
  {Aspelmeyer}}, \bibinfo {author} {\bibfnamefont {T.~J.}\ \bibnamefont
  {Kippenberg}},\ and\ \bibinfo {author} {\bibfnamefont {F.}~\bibnamefont
  {Marquardt}},\ }\bibfield  {title} {\bibinfo {title} {Cavity optomechanics},\
  }\href {https://doi.org/10.1103/RevModPhys.86.1391} {\bibfield  {journal}
  {\bibinfo  {journal} {Rev. Mod. Phys.}\ }\textbf {\bibinfo {volume} {86}},\
  \bibinfo {pages} {1391} (\bibinfo {year} {2014})},\ \Eprint
  {https://arxiv.org/abs/1303.0733} {arXiv:1303.0733 [cond-mat.mes-hall]}
  \BibitemShut {NoStop}%
\bibitem [{\citenamefont {Mancini}\ \emph {et~al.}(1998)\citenamefont
  {Mancini}, \citenamefont {Vitali},\ and\ \citenamefont
  {Tombesi}}]{Mancini1998}%
  \BibitemOpen
  \bibfield  {author} {\bibinfo {author} {\bibfnamefont {S.}~\bibnamefont
  {Mancini}}, \bibinfo {author} {\bibfnamefont {D.}~\bibnamefont {Vitali}},\
  and\ \bibinfo {author} {\bibfnamefont {P.}~\bibnamefont {Tombesi}},\
  }\bibfield  {title} {\bibinfo {title} {Optomechanical cooling of a
  macroscopic oscillator by homodyne feedback},\ }\href
  {https://doi.org/10.1103/PhysRevLett.80.688} {\bibfield  {journal} {\bibinfo
  {journal} {Phys. Rev. Lett.}\ }\textbf {\bibinfo {volume} {80}},\ \bibinfo
  {pages} {688} (\bibinfo {year} {1998})},\ \Eprint
  {https://arxiv.org/abs/9802034} {arXiv:9802034 [quant-ph]} \BibitemShut
  {NoStop}%
\bibitem [{\citenamefont {Marquardt}\ \emph {et~al.}(2007)\citenamefont
  {Marquardt}, \citenamefont {Chen}, \citenamefont {Clerk},\ and\ \citenamefont
  {Girvin}}]{Marquardt2007}%
  \BibitemOpen
  \bibfield  {author} {\bibinfo {author} {\bibfnamefont {F.}~\bibnamefont
  {Marquardt}}, \bibinfo {author} {\bibfnamefont {J.~P.}\ \bibnamefont {Chen}},
  \bibinfo {author} {\bibfnamefont {A.~A.}\ \bibnamefont {Clerk}},\ and\
  \bibinfo {author} {\bibfnamefont {S.~M.}\ \bibnamefont {Girvin}},\ }\bibfield
   {title} {\bibinfo {title} {Quantum theory of cavity-assisted sideband
  cooling of mechanical motion},\ }\href
  {https://doi.org/10.1103/PhysRevLett.99.093902} {\bibfield  {journal}
  {\bibinfo  {journal} {Phys. Rev. Lett.}\ }\textbf {\bibinfo {volume} {99}},\
  \bibinfo {pages} {093902} (\bibinfo {year} {2007})},\ \Eprint
  {https://arxiv.org/abs/0701416} {arXiv:0701416 [cond-mat]} \BibitemShut
  {NoStop}%
\bibitem [{\citenamefont {Marquardt}\ \emph {et~al.}(2008)\citenamefont
  {Marquardt}, \citenamefont {Clerk},\ and\ \citenamefont
  {Girvin}}]{Marquardt2008}%
  \BibitemOpen
  \bibfield  {author} {\bibinfo {author} {\bibfnamefont {F.}~\bibnamefont
  {Marquardt}}, \bibinfo {author} {\bibfnamefont {A.~A.}\ \bibnamefont
  {Clerk}},\ and\ \bibinfo {author} {\bibfnamefont {S.~M.}\ \bibnamefont
  {Girvin}},\ }\bibfield  {title} {\bibinfo {title} {Quantum theory of
  optomechanical cooling},\ }\href {https://doi.org/10.1080/09500340802454971}
  {\bibfield  {journal} {\bibinfo  {journal} {J. Mod. Optic.}\ }\textbf
  {\bibinfo {volume} {55}},\ \bibinfo {pages} {3329} (\bibinfo {year}
  {2008})},\ \Eprint {https://arxiv.org/abs/0803.1164} {arXiv:0803.1164
  [quant-ph]} \BibitemShut {NoStop}%
\bibitem [{\citenamefont {Yong-Chun}\ \emph {et~al.}(2013)\citenamefont
  {Yong-Chun}, \citenamefont {Yu-Wen}, \citenamefont {Wei},\ and\ \citenamefont
  {Yun-Feng}}]{Liu2013}%
  \BibitemOpen
  \bibfield  {author} {\bibinfo {author} {\bibfnamefont {L.}~\bibnamefont
  {Yong-Chun}}, \bibinfo {author} {\bibfnamefont {H.}~\bibnamefont {Yu-Wen}},
  \bibinfo {author} {\bibfnamefont {W.~C.}\ \bibnamefont {Wei}},\ and\ \bibinfo
  {author} {\bibfnamefont {X.}~\bibnamefont {Yun-Feng}},\ }\bibfield  {title}
  {\bibinfo {title} {Review of cavity optomechanical cooling},\ }\href
  {https://doi.org/10.1088/1674-1056/22/11/114213} {\bibfield  {journal}
  {\bibinfo  {journal} {Chinese Phys. B}\ }\textbf {\bibinfo {volume} {22}},\
  \bibinfo {pages} {114213} (\bibinfo {year} {2013})},\ \Eprint
  {https://arxiv.org/abs/1411.3922} {arXiv:1411.3922 [quant-ph]} \BibitemShut
  {NoStop}%
\bibitem [{\citenamefont {Weis}\ \emph {et~al.}(2010)\citenamefont {Weis},
  \citenamefont {Rivi\`{e}re}, \citenamefont {Del\'{e}glise}, \citenamefont
  {Gavartin}, \citenamefont {Arcizet}, \citenamefont {Schliesser},\ and\
  \citenamefont {Kippenberg}}]{Weis2010}%
  \BibitemOpen
  \bibfield  {author} {\bibinfo {author} {\bibfnamefont {S.}~\bibnamefont
  {Weis}}, \bibinfo {author} {\bibfnamefont {R.}~\bibnamefont {Rivi\`{e}re}},
  \bibinfo {author} {\bibfnamefont {S.}~\bibnamefont {Del\'{e}glise}}, \bibinfo
  {author} {\bibfnamefont {E.}~\bibnamefont {Gavartin}}, \bibinfo {author}
  {\bibfnamefont {O.}~\bibnamefont {Arcizet}}, \bibinfo {author} {\bibfnamefont
  {A.}~\bibnamefont {Schliesser}},\ and\ \bibinfo {author} {\bibfnamefont
  {T.~J.}\ \bibnamefont {Kippenberg}},\ }\bibfield  {title} {\bibinfo {title}
  {Optomechanically induced transparency},\ }\href
  {https://doi.org/10.1126/science.1195596} {\bibfield  {journal} {\bibinfo
  {journal} {Science}\ }\textbf {\bibinfo {volume} {330}},\ \bibinfo {pages}
  {1520} (\bibinfo {year} {2010})},\ \Eprint {https://arxiv.org/abs/1007.0565}
  {arXiv:1007.0565 [quant-ph]} \BibitemShut {NoStop}%
\bibitem [{\citenamefont {Karuza}\ \emph {et~al.}(2013)\citenamefont {Karuza},
  \citenamefont {Biancofiore}, \citenamefont {Bawaj}, \citenamefont
  {Molinelli}, \citenamefont {Galassi}, \citenamefont {Natali}, \citenamefont
  {Tombesi}, \citenamefont {{Di Giuseppe}},\ and\ \citenamefont
  {Vitali}}]{Karuza2013}%
  \BibitemOpen
  \bibfield  {author} {\bibinfo {author} {\bibfnamefont {M.}~\bibnamefont
  {Karuza}}, \bibinfo {author} {\bibfnamefont {C.}~\bibnamefont {Biancofiore}},
  \bibinfo {author} {\bibfnamefont {M.}~\bibnamefont {Bawaj}}, \bibinfo
  {author} {\bibfnamefont {C.}~\bibnamefont {Molinelli}}, \bibinfo {author}
  {\bibfnamefont {M.}~\bibnamefont {Galassi}}, \bibinfo {author} {\bibfnamefont
  {R.}~\bibnamefont {Natali}}, \bibinfo {author} {\bibfnamefont
  {P.}~\bibnamefont {Tombesi}}, \bibinfo {author} {\bibfnamefont
  {G.}~\bibnamefont {{Di Giuseppe}}},\ and\ \bibinfo {author} {\bibfnamefont
  {D.}~\bibnamefont {Vitali}},\ }\bibfield  {title} {\bibinfo {title}
  {Optomechanically induced transparency in a membrane-in-the-middle setup at
  room temperature},\ }\href {https://doi.org/10.1103/PhysRevA.88.013804}
  {\bibfield  {journal} {\bibinfo  {journal} {Phys. Rev. A}\ }\textbf {\bibinfo
  {volume} {88}},\ \bibinfo {pages} {013804} (\bibinfo {year} {2013})},\
  \Eprint {https://arxiv.org/abs/1209.1352} {arXiv:1209.1352 [quant-ph]}
  \BibitemShut {NoStop}%
\bibitem [{\citenamefont {Frank}\ \emph {et~al.}(2010)\citenamefont {Frank},
  \citenamefont {Deotare}, \citenamefont {Mc{C}utcheon},\ and\ \citenamefont
  {Loncar}}]{Frank2010}%
  \BibitemOpen
  \bibfield  {author} {\bibinfo {author} {\bibfnamefont {I.~W.}\ \bibnamefont
  {Frank}}, \bibinfo {author} {\bibfnamefont {P.~B.}\ \bibnamefont {Deotare}},
  \bibinfo {author} {\bibfnamefont {M.~W.}\ \bibnamefont {Mc{C}utcheon}},\ and\
  \bibinfo {author} {\bibfnamefont {M.}~\bibnamefont {Loncar}},\ }\bibfield
  {title} {\bibinfo {title} {Programmable photonic crystal nanobeam cavities},\
  }\href {https://doi.org/htpps://doi.org/10.1364/OE.18.008705} {\bibfield
  {journal} {\bibinfo  {journal} {Opt. Express}\ }\textbf {\bibinfo {volume}
  {18}},\ \bibinfo {pages} {8705} (\bibinfo {year} {2010})}\BibitemShut
  {NoStop}%
\bibitem [{\citenamefont {Qiao}\ \emph {et~al.}(2018)\citenamefont {Qiao},
  \citenamefont {Xia}, \citenamefont {Lee},\ and\ \citenamefont
  {Zhou}}]{Qiao2018}%
  \BibitemOpen
  \bibfield  {author} {\bibinfo {author} {\bibfnamefont {Q.}~\bibnamefont
  {Qiao}}, \bibinfo {author} {\bibfnamefont {J.}~\bibnamefont {Xia}}, \bibinfo
  {author} {\bibfnamefont {C.}~\bibnamefont {Lee}},\ and\ \bibinfo {author}
  {\bibfnamefont {G.}~\bibnamefont {Zhou}},\ }\bibfield  {title} {\bibinfo
  {title} {Applications of photonic crystal nanobeam cavities for sensing},\
  }\href {https://doi.org/10.3390/mi9110541} {\bibfield  {journal} {\bibinfo
  {journal} {Micromachines}\ }\textbf {\bibinfo {volume} {9}},\ \bibinfo
  {pages} {541} (\bibinfo {year} {2018})}\BibitemShut {NoStop}%
\bibitem [{\citenamefont {Zhou}\ \emph {et~al.}(2014)\citenamefont {Zhou},
  \citenamefont {Tian}, \citenamefont {Yang}, \citenamefont {Liu},
  \citenamefont {Huang},\ and\ \citenamefont {Ji}}]{Zhou2014}%
  \BibitemOpen
  \bibfield  {author} {\bibinfo {author} {\bibfnamefont {J.}~\bibnamefont
  {Zhou}}, \bibinfo {author} {\bibfnamefont {H.}~\bibnamefont {Tian}}, \bibinfo
  {author} {\bibfnamefont {D.}~\bibnamefont {Yang}}, \bibinfo {author}
  {\bibfnamefont {Q.}~\bibnamefont {Liu}}, \bibinfo {author} {\bibfnamefont
  {L.}~\bibnamefont {Huang}},\ and\ \bibinfo {author} {\bibfnamefont
  {Y.}~\bibnamefont {Ji}},\ }\bibfield  {title} {\bibinfo {title} {Refractive
  index sensing utilizing parallel tapered nano-slotted photonic crystal
  nano-beam cavities},\ }\href {https://doi.org/10.1364/JOSAB.31.001746}
  {\bibfield  {journal} {\bibinfo  {journal} {J. Opt. Soc. Am. B}\ }\textbf
  {\bibinfo {volume} {31}},\ \bibinfo {pages} {1746} (\bibinfo {year}
  {2014})}\BibitemShut {NoStop}%
\bibitem [{\citenamefont {Yang}\ \emph {et~al.}(2020)\citenamefont {Yang},
  \citenamefont {Duan}, \citenamefont {Liu}, \citenamefont {Wang},
  \citenamefont {Li},\ and\ \citenamefont {Ji}}]{Yang2020}%
  \BibitemOpen
  \bibfield  {author} {\bibinfo {author} {\bibfnamefont {D.-Q.}\ \bibnamefont
  {Yang}}, \bibinfo {author} {\bibfnamefont {B.}~\bibnamefont {Duan}}, \bibinfo
  {author} {\bibfnamefont {X.}~\bibnamefont {Liu}}, \bibinfo {author}
  {\bibfnamefont {A.-Q.}\ \bibnamefont {Wang}}, \bibinfo {author}
  {\bibfnamefont {X.-G.}\ \bibnamefont {Li}},\ and\ \bibinfo {author}
  {\bibfnamefont {Y.-F.}\ \bibnamefont {Ji}},\ }\bibfield  {title} {\bibinfo
  {title} {Photonic crystal nanobeam cavities for nanoscale optical sensing: A
  review},\ }\href {https://doi.org/10.3390/mi11010072} {\bibfield  {journal}
  {\bibinfo  {journal} {Micromachines}\ }\textbf {\bibinfo {volume} {11}},\
  \bibinfo {pages} {72} (\bibinfo {year} {2020})}\BibitemShut {NoStop}%
\bibitem [{\citenamefont {Pietikainen}\ \emph {et~al.}(2022)\citenamefont
  {Pietikainen}, \citenamefont {Cernotik}, \citenamefont {Puri},\ and\
  \citenamefont {Filip}}]{Pietikainen2022}%
  \BibitemOpen
  \bibfield  {author} {\bibinfo {author} {\bibfnamefont {I.}~\bibnamefont
  {Pietikainen}}, \bibinfo {author} {\bibfnamefont {O.}~\bibnamefont
  {Cernotik}}, \bibinfo {author} {\bibfnamefont {S.}~\bibnamefont {Puri}},\
  and\ \bibinfo {author} {\bibfnamefont {R.}~\bibnamefont {Filip}},\ }\bibfield
   {title} {\bibinfo {title} {Controlled beam splitter gate transparent to
  dominant ancilla errors},\ }\href {https://doi.org/10.1088/2058-9565/ac760a}
  {\bibfield  {journal} {\bibinfo  {journal} {Quantum Sci. Technol.}\ }\textbf
  {\bibinfo {volume} {7}},\ \bibinfo {pages} {035025} (\bibinfo {year}
  {2022})}\BibitemShut {NoStop}%
\bibitem [{\citenamefont {Gu}\ \emph {et~al.}(2018)\citenamefont {Gu},
  \citenamefont {Yi}, \citenamefont {Sun},\ and\ \citenamefont {Yan}}]{Gu2018}%
  \BibitemOpen
  \bibfield  {author} {\bibinfo {author} {\bibfnamefont {W.~J.}\ \bibnamefont
  {Gu}}, \bibinfo {author} {\bibfnamefont {Z.}~\bibnamefont {Yi}}, \bibinfo
  {author} {\bibfnamefont {L.~H.}\ \bibnamefont {Sun}},\ and\ \bibinfo {author}
  {\bibfnamefont {Y.}~\bibnamefont {Yan}},\ }\bibfield  {title} {\bibinfo
  {title} {Generation of mechanical squeezing and entanglement via mechanical
  modulations},\ }\href {https://doi.org/10.1364/OE.26.030773} {\bibfield
  {journal} {\bibinfo  {journal} {Opt. Express}\ }\textbf {\bibinfo {volume}
  {26}},\ \bibinfo {pages} {30773} (\bibinfo {year} {2018})}\BibitemShut
  {NoStop}%
\bibitem [{\citenamefont {Li}\ \emph {et~al.}(2021)\citenamefont {Li},
  \citenamefont {Ou}, \citenamefont {Lei},\ and\ \citenamefont {Liu}}]{Li2021}%
  \BibitemOpen
  \bibfield  {author} {\bibinfo {author} {\bibfnamefont {B.-B.}\ \bibnamefont
  {Li}}, \bibinfo {author} {\bibfnamefont {L.}~\bibnamefont {Ou}}, \bibinfo
  {author} {\bibfnamefont {Y.}~\bibnamefont {Lei}},\ and\ \bibinfo {author}
  {\bibfnamefont {Y.-C.}\ \bibnamefont {Liu}},\ }\bibfield  {title} {\bibinfo
  {title} {Cavity optomechanical sensing},\ }\href
  {https://doi.org/10.1515/nanoph-2021-0256} {\bibfield  {journal} {\bibinfo
  {journal} {Nanophotonics}\ }\textbf {\bibinfo {volume} {10}},\ \bibinfo
  {pages} {2799} (\bibinfo {year} {2021})}\BibitemShut {NoStop}%
\bibitem [{\citenamefont {Guha}\ \emph {et~al.}(2021)\citenamefont {Guha},
  \citenamefont {Wu}, \citenamefont {{Dong Song}}, \citenamefont {Balram},\
  and\ \citenamefont {Srinivasan}}]{Guha2021}%
  \BibitemOpen
  \bibfield  {author} {\bibinfo {author} {\bibfnamefont {B.}~\bibnamefont
  {Guha}}, \bibinfo {author} {\bibfnamefont {M.}~\bibnamefont {Wu}}, \bibinfo
  {author} {\bibfnamefont {J.}~\bibnamefont {{Dong Song}}}, \bibinfo {author}
  {\bibfnamefont {K.~C.}\ \bibnamefont {Balram}},\ and\ \bibinfo {author}
  {\bibfnamefont {K.}~\bibnamefont {Srinivasan}},\ }\bibfield  {title}
  {\bibinfo {title} {Piezo-optomechanical actuation of nanobeam resonators for
  microwave-to-optical transduction},\ }in\ \href@noop {} {\emph {\bibinfo
  {booktitle} {2021 Conference on Lasers and Electro-Optics ({CLEO})}}}\
  (\bibinfo {year} {2021})\ p.~\bibinfo {pages} {1}\BibitemShut {NoStop}%
\bibitem [{\citenamefont {Balram}\ and\ \citenamefont
  {Srinivasan}(2022)}]{Balram2022}%
  \BibitemOpen
  \bibfield  {author} {\bibinfo {author} {\bibfnamefont {K.~C.}\ \bibnamefont
  {Balram}}\ and\ \bibinfo {author} {\bibfnamefont {K.}~\bibnamefont
  {Srinivasan}},\ }\bibfield  {title} {\bibinfo {title} {Piezoelectric
  optomechanical approaches for efficient quantum microwave-to-optical signal
  transduction: The need for co-design},\ }\href
  {https://doi.org/10.1002/qute.202100095} {\bibfield  {journal} {\bibinfo
  {journal} {Adv. Quantum Technol.}\ ,\ \bibinfo {pages} {2100095}} (\bibinfo
  {year} {2022})},\ \Eprint {https://arxiv.org/abs/2108.11797}
  {arXiv:2108.11797 [physics.optics]} \BibitemShut {NoStop}%
\bibitem [{\citenamefont {Braginsky}\ and\ \citenamefont
  {Manukin}(1967)}]{Braginsky1967a}%
  \BibitemOpen
  \bibfield  {author} {\bibinfo {author} {\bibfnamefont {V.~B.}\ \bibnamefont
  {Braginsky}}\ and\ \bibinfo {author} {\bibfnamefont {A.~B.}\ \bibnamefont
  {Manukin}},\ }\bibfield  {title} {\bibinfo {title} {Ponderomotive effects of
  electromagnetic radiation},\ }\href
  {http://www.jetp.ras.ru/cgi-bin/dn/e_028_02_0301.pdf} {\bibfield  {journal}
  {\bibinfo  {journal} {Sov. Phys. J. Exp. Theor. Phys.}\ }\textbf {\bibinfo
  {volume} {25}},\ \bibinfo {pages} {653} (\bibinfo {year} {1967})}\BibitemShut
  {NoStop}%
\bibitem [{\citenamefont {Braginsky}\ \emph {et~al.}(1970)\citenamefont
  {Braginsky}, \citenamefont {Manukin},\ and\ \citenamefont
  {Tikhonov}}]{Braginsky1970}%
  \BibitemOpen
  \bibfield  {author} {\bibinfo {author} {\bibfnamefont {V.~B.}\ \bibnamefont
  {Braginsky}}, \bibinfo {author} {\bibfnamefont {A.~B.}\ \bibnamefont
  {Manukin}},\ and\ \bibinfo {author} {\bibfnamefont {M.~Y.}\ \bibnamefont
  {Tikhonov}},\ }\bibfield  {title} {\bibinfo {title} {Investigation of
  dissipative ponderomotive effects of electromagnetic radiation},\ }\href
  {http://jetp.ras.ru/cgi-bin/dn/e_031_05_0829.pdf} {\bibfield  {journal}
  {\bibinfo  {journal} {Sov. Phys. J. Exp. Theor. Phys.}\ }\textbf {\bibinfo
  {volume} {31}},\ \bibinfo {pages} {829} (\bibinfo {year} {1970})}\BibitemShut
  {NoStop}%
\bibitem [{\citenamefont {Xu}\ \emph {et~al.}(2015)\citenamefont {Xu},
  \citenamefont {x.~Liu}, \citenamefont {Sun},\ and\ \citenamefont
  {Li}}]{Xu2015}%
  \BibitemOpen
  \bibfield  {author} {\bibinfo {author} {\bibfnamefont {X.-W.}\ \bibnamefont
  {Xu}}, \bibinfo {author} {\bibfnamefont {Y.}~\bibnamefont {x.~Liu}}, \bibinfo
  {author} {\bibfnamefont {C.-P.}\ \bibnamefont {Sun}},\ and\ \bibinfo {author}
  {\bibfnamefont {Y.}~\bibnamefont {Li}},\ }\bibfield  {title} {\bibinfo
  {title} {Mechanical $\mathcal{PT}$ symmetry in coupled optomechanical
  systems},\ }\href {https://doi.org/10.1103/PhysRevA.92.013852} {\bibfield
  {journal} {\bibinfo  {journal} {Phys. Rev. A}\ }\textbf {\bibinfo {volume}
  {92}},\ \bibinfo {pages} {013852} (\bibinfo {year} {2015})},\ \Eprint
  {https://arxiv.org/abs/1402.7222} {arXiv:1402.7222 [quant-ph]} \BibitemShut
  {NoStop}%
\bibitem [{\citenamefont {Qiu}\ \emph {et~al.}(2020)\citenamefont {Qiu},
  \citenamefont {Shomroni}, \citenamefont {Seidler},\ and\ \citenamefont
  {Kippenberg}}]{Qiu2020}%
  \BibitemOpen
  \bibfield  {author} {\bibinfo {author} {\bibfnamefont {L.}~\bibnamefont
  {Qiu}}, \bibinfo {author} {\bibfnamefont {I.}~\bibnamefont {Shomroni}},
  \bibinfo {author} {\bibfnamefont {P.}~\bibnamefont {Seidler}},\ and\ \bibinfo
  {author} {\bibfnamefont {T.~J.}\ \bibnamefont {Kippenberg}},\ }\bibfield
  {title} {\bibinfo {title} {Laser cooling of a nanomechanical oscillator to
  its zero-point energy},\ }\href
  {https://doi.org/10.1103/PhysRevLett.124.173601} {\bibfield  {journal}
  {\bibinfo  {journal} {Phys. Rev. Lett.}\ }\textbf {\bibinfo {volume} {124}},\
  \bibinfo {pages} {173601} (\bibinfo {year} {2020})},\ \Eprint
  {https://arxiv.org/abs/1903.10242} {arXiv:1903.10242 [quant-ph]} \BibitemShut
  {NoStop}%
\bibitem [{\citenamefont {Xue}\ \emph {et~al.}(2007)\citenamefont {Xue},
  \citenamefont {Liu}, \citenamefont {Sun},\ and\ \citenamefont
  {Nori}}]{Xue2007}%
  \BibitemOpen
  \bibfield  {author} {\bibinfo {author} {\bibfnamefont {F.}~\bibnamefont
  {Xue}}, \bibinfo {author} {\bibfnamefont {Y.~X.}\ \bibnamefont {Liu}},
  \bibinfo {author} {\bibfnamefont {C.~P.}\ \bibnamefont {Sun}},\ and\ \bibinfo
  {author} {\bibfnamefont {F.}~\bibnamefont {Nori}},\ }\bibfield  {title}
  {\bibinfo {title} {Two-mode squeezed states and entangled states of two
  mechanical resonators},\ }\href {https://doi.org/10.1103/PhysRevB.76.064305}
  {\bibfield  {journal} {\bibinfo  {journal} {Phys. Rev. B}\ }\textbf {\bibinfo
  {volume} {76}},\ \bibinfo {pages} {064305} (\bibinfo {year} {2007})},\
  \Eprint {https://arxiv.org/abs/0701209} {arXiv:0701209 [quant-ph]}
  \BibitemShut {NoStop}%
\bibitem [{\citenamefont {Tan}\ \emph {et~al.}(2013)\citenamefont {Tan},
  \citenamefont {Li},\ and\ \citenamefont {Meystre}}]{Tan2013}%
  \BibitemOpen
  \bibfield  {author} {\bibinfo {author} {\bibfnamefont {H.}~\bibnamefont
  {Tan}}, \bibinfo {author} {\bibfnamefont {G.}~\bibnamefont {Li}},\ and\
  \bibinfo {author} {\bibfnamefont {P.}~\bibnamefont {Meystre}},\ }\bibfield
  {title} {\bibinfo {title} {Dissipation-driven two mode mechanical squeezing
  states in optomechanical systems},\ }\href
  {https://doi.org/10.1103/PhysRevA.87.033829} {\bibfield  {journal} {\bibinfo
  {journal} {Phys. Rev. A}\ }\textbf {\bibinfo {volume} {87}},\ \bibinfo
  {pages} {033829} (\bibinfo {year} {2013})},\ \Eprint
  {https://arxiv.org/abs/1301.5698} {arXiv:1301.5698 [quant-ph]} \BibitemShut
  {NoStop}%
\bibitem [{\citenamefont {Woolley}\ and\ \citenamefont
  {Clerk}(2014)}]{Woolley2014}%
  \BibitemOpen
  \bibfield  {author} {\bibinfo {author} {\bibfnamefont {M.~J.}\ \bibnamefont
  {Woolley}}\ and\ \bibinfo {author} {\bibfnamefont {A.~A.}\ \bibnamefont
  {Clerk}},\ }\bibfield  {title} {\bibinfo {title} {Two-mode squeezing states
  in cavity optomechanics via engineering of a single reservoir},\ }\href
  {https://doi.org/10.1103/PhysRevA.89.063805} {\bibfield  {journal} {\bibinfo
  {journal} {Phys. Rev. A}\ }\textbf {\bibinfo {volume} {89}},\ \bibinfo
  {pages} {063805} (\bibinfo {year} {2014})},\ \Eprint
  {https://arxiv.org/abs/1404.2672} {arXiv:1404.2672 [quant-ph]} \BibitemShut
  {NoStop}%
\bibitem [{\citenamefont {Pontin}\ \emph {et~al.}(2016)\citenamefont {Pontin},
  \citenamefont {Bonaldi}, \citenamefont {Borrielli}, \citenamefont {Marconi},
  \citenamefont {Marino}, \citenamefont {Pandraud}, \citenamefont {Prodi},
  \citenamefont {Sarro}, \citenamefont {Serra},\ and\ \citenamefont
  {Marin}}]{Pontin2016}%
  \BibitemOpen
  \bibfield  {author} {\bibinfo {author} {\bibfnamefont {A.}~\bibnamefont
  {Pontin}}, \bibinfo {author} {\bibfnamefont {M.}~\bibnamefont {Bonaldi}},
  \bibinfo {author} {\bibfnamefont {A.}~\bibnamefont {Borrielli}}, \bibinfo
  {author} {\bibfnamefont {L.}~\bibnamefont {Marconi}}, \bibinfo {author}
  {\bibfnamefont {F.}~\bibnamefont {Marino}}, \bibinfo {author} {\bibfnamefont
  {G.}~\bibnamefont {Pandraud}}, \bibinfo {author} {\bibfnamefont {G.~A.}\
  \bibnamefont {Prodi}}, \bibinfo {author} {\bibfnamefont {P.~M.}\ \bibnamefont
  {Sarro}}, \bibinfo {author} {\bibfnamefont {E.}~\bibnamefont {Serra}},\ and\
  \bibinfo {author} {\bibfnamefont {F.}~\bibnamefont {Marin}},\ }\bibfield
  {title} {\bibinfo {title} {Dynamical two-mode squeezing of thermal
  fluctuations in a cavity optomechanical system},\ }\href
  {https://doi.org/10.1103/PhysRevLett.116.103601} {\bibfield  {journal}
  {\bibinfo  {journal} {Phys. Rev. Lett.}\ }\textbf {\bibinfo {volume} {116}},\
  \bibinfo {pages} {103601} (\bibinfo {year} {2016})},\ \Eprint
  {https://arxiv.org/abs/1509.02723} {arXiv:1509.02723 [quant-ph]} \BibitemShut
  {NoStop}%
\bibitem [{\citenamefont {Mahbood}\ \emph {et~al.}(2014)\citenamefont
  {Mahbood}, \citenamefont {Okamoto}, \citenamefont {Onomitsu},\ and\
  \citenamefont {Yamaguchi}}]{Mahbood2014}%
  \BibitemOpen
  \bibfield  {author} {\bibinfo {author} {\bibfnamefont {I.}~\bibnamefont
  {Mahbood}}, \bibinfo {author} {\bibfnamefont {H.}~\bibnamefont {Okamoto}},
  \bibinfo {author} {\bibfnamefont {K.}~\bibnamefont {Onomitsu}},\ and\
  \bibinfo {author} {\bibfnamefont {H.}~\bibnamefont {Yamaguchi}},\ }\bibfield
  {title} {\bibinfo {title} {Two-mode thermal-noise squeezing in an
  electromechanical resonator},\ }\href
  {https://doi.org/10.1103/PhysRevLett.113.167203} {\bibfield  {journal}
  {\bibinfo  {journal} {Phys. Rev. Lett}\ }\textbf {\bibinfo {volume} {113}},\
  \bibinfo {pages} {167203} (\bibinfo {year} {2014})},\ \Eprint
  {https://arxiv.org/abs/1405.5270} {arXiv:1405.5270 [cond-mat.mes-hall]}
  \BibitemShut {NoStop}%
\bibitem [{\citenamefont {Patil}\ \emph {et~al.}(2015)\citenamefont {Patil},
  \citenamefont {Chakram}, \citenamefont {Chang},\ and\ \citenamefont
  {Vengalattore}}]{Patil2015}%
  \BibitemOpen
  \bibfield  {author} {\bibinfo {author} {\bibfnamefont {Y.~S.}\ \bibnamefont
  {Patil}}, \bibinfo {author} {\bibfnamefont {S.}~\bibnamefont {Chakram}},
  \bibinfo {author} {\bibfnamefont {L.}~\bibnamefont {Chang}},\ and\ \bibinfo
  {author} {\bibfnamefont {M.}~\bibnamefont {Vengalattore}},\ }\bibfield
  {title} {\bibinfo {title} {Thermomechanical two-mode squeezing in an
  ultrahigh-{Q} membrane resonator},\ }\href
  {https://doi.org/10.1103/PhysRevLett.115.017202} {\bibfield  {journal}
  {\bibinfo  {journal} {Phys. Rev. Lett.}\ }\textbf {\bibinfo {volume} {115}},\
  \bibinfo {pages} {017202} (\bibinfo {year} {2015})},\ \Eprint
  {https://arxiv.org/abs/1410.7109} {arXiv:1410.7109 [quant-ph]} \BibitemShut
  {NoStop}%
\bibitem [{\citenamefont {Shakeri}\ \emph {et~al.}(2016)\citenamefont
  {Shakeri}, \citenamefont {Mahmoudi}, \citenamefont {Zandi},\ and\
  \citenamefont {Bahrampour}}]{Shakeri2016}%
  \BibitemOpen
  \bibfield  {author} {\bibinfo {author} {\bibfnamefont {S.}~\bibnamefont
  {Shakeri}}, \bibinfo {author} {\bibfnamefont {Z.}~\bibnamefont {Mahmoudi}},
  \bibinfo {author} {\bibfnamefont {M.~H.}\ \bibnamefont {Zandi}},\ and\
  \bibinfo {author} {\bibfnamefont {A.~R.}\ \bibnamefont {Bahrampour}},\
  }\bibfield  {title} {\bibinfo {title} {Two mode mechanical non-{G}aussian
  squeezed number state in a two-membrane optomechanical system},\ }\href
  {https://doi.org/10.1016/j.optcom.2016.02.063} {\bibfield  {journal}
  {\bibinfo  {journal} {Opt. Commun.}\ }\textbf {\bibinfo {volume} {370}},\
  \bibinfo {pages} {55} (\bibinfo {year} {2016})}\BibitemShut {NoStop}%
\bibitem [{\citenamefont {Martini}\ and\ \citenamefont
  {Sciarrino}(2005)}]{DeMartini2005}%
  \BibitemOpen
  \bibfield  {author} {\bibinfo {author} {\bibfnamefont {F.~D.}\ \bibnamefont
  {Martini}}\ and\ \bibinfo {author} {\bibfnamefont {F.}~\bibnamefont
  {Sciarrino}},\ }\bibfield  {title} {\bibinfo {title} {Review on non-linear
  parametric processes in quantum information},\ }\href
  {https://doi.org/10.1016/j.pquantelec.2005.08.001} {\bibfield  {journal}
  {\bibinfo  {journal} {Prog. Quant. Electron.}\ }\textbf {\bibinfo {volume}
  {29}},\ \bibinfo {pages} {165} (\bibinfo {year} {2005})}\BibitemShut
  {NoStop}%
\bibitem [{\citenamefont {Piergentili}\ \emph {et~al.}(2021)\citenamefont
  {Piergentili}, \citenamefont {Li}, \citenamefont {Natali}, \citenamefont
  {Malossi}, \citenamefont {Vitali},\ and\ \citenamefont {{Di
  Giuseppe}}}]{Piergentili2021}%
  \BibitemOpen
  \bibfield  {author} {\bibinfo {author} {\bibfnamefont {P.}~\bibnamefont
  {Piergentili}}, \bibinfo {author} {\bibfnamefont {W.}~\bibnamefont {Li}},
  \bibinfo {author} {\bibfnamefont {R.}~\bibnamefont {Natali}}, \bibinfo
  {author} {\bibfnamefont {N.}~\bibnamefont {Malossi}}, \bibinfo {author}
  {\bibfnamefont {D.}~\bibnamefont {Vitali}},\ and\ \bibinfo {author}
  {\bibfnamefont {G.}~\bibnamefont {{Di Giuseppe}}},\ }\bibfield  {title}
  {\bibinfo {title} {Two-membrane cavity optomechanics: non-linear dynamics},\
  }\href {https://doi.org/10.1088/1367-2630/abdd6a} {\bibfield  {journal}
  {\bibinfo  {journal} {New J. Phys.}\ }\textbf {\bibinfo {volume} {23}},\
  \bibinfo {pages} {1367} (\bibinfo {year} {2021})},\ \Eprint
  {https://arxiv.org/abs/2009.04694} {arXiv:2009.04694 [quant-ph]} \BibitemShut
  {NoStop}%
\bibitem [{\citenamefont {Stannigel}\ \emph {et~al.}(2010)\citenamefont
  {Stannigel}, \citenamefont {Rabl}, \citenamefont {S\o{}rensen}, \citenamefont
  {Zoller},\ and\ \citenamefont {Lukin}}]{Stannigel2010}%
  \BibitemOpen
  \bibfield  {author} {\bibinfo {author} {\bibfnamefont {K.}~\bibnamefont
  {Stannigel}}, \bibinfo {author} {\bibfnamefont {P.}~\bibnamefont {Rabl}},
  \bibinfo {author} {\bibfnamefont {A.~S.}\ \bibnamefont {S\o{}rensen}},
  \bibinfo {author} {\bibfnamefont {P.}~\bibnamefont {Zoller}},\ and\ \bibinfo
  {author} {\bibfnamefont {M.~D.}\ \bibnamefont {Lukin}},\ }\bibfield  {title}
  {\bibinfo {title} {Optomechanical transducers for long-distance quantum
  communication},\ }\href {https://doi.org/10.1103/PhysRevLett.105.220501}
  {\bibfield  {journal} {\bibinfo  {journal} {Phys. Rev. Lett.}\ }\textbf
  {\bibinfo {volume} {105}},\ \bibinfo {pages} {220501} (\bibinfo {year}
  {2010})},\ \Eprint {https://arxiv.org/abs/1006.4361} {arXiv:1006.4361
  [quant-ph]} \BibitemShut {NoStop}%
\bibitem [{\citenamefont {Stannigel}\ \emph {et~al.}(2011)\citenamefont
  {Stannigel}, \citenamefont {Rabl}, \citenamefont {S\o{}rensen}, \citenamefont
  {Lukin},\ and\ \citenamefont {Zoller}}]{Stannigel2011}%
  \BibitemOpen
  \bibfield  {author} {\bibinfo {author} {\bibfnamefont {K.}~\bibnamefont
  {Stannigel}}, \bibinfo {author} {\bibfnamefont {P.}~\bibnamefont {Rabl}},
  \bibinfo {author} {\bibfnamefont {A.~S.}\ \bibnamefont {S\o{}rensen}},
  \bibinfo {author} {\bibfnamefont {M.~D.}\ \bibnamefont {Lukin}},\ and\
  \bibinfo {author} {\bibfnamefont {P.}~\bibnamefont {Zoller}},\ }\bibfield
  {title} {\bibinfo {title} {Optomechanical transducers for quantum-information
  processing},\ }\href {https://doi.org/10.1103/PhysRevA.84.042341} {\bibfield
  {journal} {\bibinfo  {journal} {Phys. Rev. A}\ }\textbf {\bibinfo {volume}
  {84}},\ \bibinfo {pages} {042341} (\bibinfo {year} {2011})},\ \Eprint
  {https://arxiv.org/abs/1106.5394} {arXiv:1106.5394 [quant-ph]} \BibitemShut
  {NoStop}%
\bibitem [{\citenamefont {Cole}\ and\ \citenamefont
  {Aspelmeyer}(2011)}]{Cole2011}%
  \BibitemOpen
  \bibfield  {author} {\bibinfo {author} {\bibfnamefont {G.~D.}\ \bibnamefont
  {Cole}}\ and\ \bibinfo {author} {\bibfnamefont {M.}~\bibnamefont
  {Aspelmeyer}},\ }\bibfield  {title} {\bibinfo {title} {Mechanical memory sees
  the light},\ }\href {https://doi.org/10.1038/nnano.2011.199} {\bibfield
  {journal} {\bibinfo  {journal} {Nat. Nanotechnol.}\ }\textbf {\bibinfo
  {volume} {6}},\ \bibinfo {pages} {690} (\bibinfo {year} {2011})}\BibitemShut
  {NoStop}%
\bibitem [{\citenamefont {Fiaschi}\ \emph {et~al.}(2021)\citenamefont
  {Fiaschi}, \citenamefont {Hensen}, \citenamefont {Wallucks}, \citenamefont
  {Benevides}, \citenamefont {Li}, \citenamefont {{Mayer Alegre}},\ and\
  \citenamefont {Gr\"{o}blacher}}]{Fiaschi2021}%
  \BibitemOpen
  \bibfield  {author} {\bibinfo {author} {\bibfnamefont {N.}~\bibnamefont
  {Fiaschi}}, \bibinfo {author} {\bibfnamefont {B.}~\bibnamefont {Hensen}},
  \bibinfo {author} {\bibfnamefont {A.}~\bibnamefont {Wallucks}}, \bibinfo
  {author} {\bibfnamefont {R.}~\bibnamefont {Benevides}}, \bibinfo {author}
  {\bibfnamefont {J.}~\bibnamefont {Li}}, \bibinfo {author} {\bibfnamefont
  {T.~P.}\ \bibnamefont {{Mayer Alegre}}},\ and\ \bibinfo {author}
  {\bibfnamefont {S.}~\bibnamefont {Gr\"{o}blacher}},\ }\bibfield  {title}
  {\bibinfo {title} {Optomechanical quantum teleportation},\ }\href
  {https://doi.org/10.1038/s41566-021-00866-z} {\bibfield  {journal} {\bibinfo
  {journal} {Nat. Photonics}\ }\textbf {\bibinfo {volume} {15}},\ \bibinfo
  {pages} {817} (\bibinfo {year} {2021})},\ \Eprint
  {https://arxiv.org/abs/1106.5394} {arXiv:1106.5394 [quant-ph]} \BibitemShut
  {NoStop}%
\bibitem [{\citenamefont {Eichenfield}\ \emph {et~al.}(2009)\citenamefont
  {Eichenfield}, \citenamefont {Chan}, \citenamefont {Safavi-Naeini},
  \citenamefont {Vahala},\ and\ \citenamefont {Painter}}]{Eichenfield2009}%
  \BibitemOpen
  \bibfield  {author} {\bibinfo {author} {\bibfnamefont {M.}~\bibnamefont
  {Eichenfield}}, \bibinfo {author} {\bibfnamefont {J.}~\bibnamefont {Chan}},
  \bibinfo {author} {\bibfnamefont {A.~H.}\ \bibnamefont {Safavi-Naeini}},
  \bibinfo {author} {\bibfnamefont {K.~J.}\ \bibnamefont {Vahala}},\ and\
  \bibinfo {author} {\bibfnamefont {O.}~\bibnamefont {Painter}},\ }\bibfield
  {title} {\bibinfo {title} {Modeling dispersive coupling and losses of
  localized optical and mechanical modes in optomechanical crystals},\ }\href
  {https://doi.org/10.1364/OE.17.020078} {\bibfield  {journal} {\bibinfo
  {journal} {Opt. Express}\ }\textbf {\bibinfo {volume} {17}},\ \bibinfo
  {pages} {20078} (\bibinfo {year} {2009})},\ \Eprint
  {https://arxiv.org/abs/0908.0025} {arXiv:0908.0025 [physics.optics]}
  \BibitemShut {NoStop}%
\bibitem [{\citenamefont {Chan}\ \emph {et~al.}(2011)\citenamefont {Chan},
  \citenamefont {{Mayer Alegre}}, \citenamefont {Safavi-Naeini}, \citenamefont
  {Hill}, \citenamefont {Krause}, \citenamefont {Groeblacher}, \citenamefont
  {Aspelmeyer},\ and\ \citenamefont {Painter}}]{Chan2011}%
  \BibitemOpen
  \bibfield  {author} {\bibinfo {author} {\bibfnamefont {J.}~\bibnamefont
  {Chan}}, \bibinfo {author} {\bibfnamefont {T.~P.}\ \bibnamefont {{Mayer
  Alegre}}}, \bibinfo {author} {\bibfnamefont {A.~H.}\ \bibnamefont
  {Safavi-Naeini}}, \bibinfo {author} {\bibfnamefont {J.~T.}\ \bibnamefont
  {Hill}}, \bibinfo {author} {\bibfnamefont {A.}~\bibnamefont {Krause}},
  \bibinfo {author} {\bibfnamefont {S.}~\bibnamefont {Groeblacher}}, \bibinfo
  {author} {\bibfnamefont {M.}~\bibnamefont {Aspelmeyer}},\ and\ \bibinfo
  {author} {\bibfnamefont {O.}~\bibnamefont {Painter}},\ }\bibfield  {title}
  {\bibinfo {title} {Laser cooling of a nanomechanical oscillator into its
  quantum ground state},\ }\href {https://doi.org/10.1038/nature10461}
  {\bibfield  {journal} {\bibinfo  {journal} {Nature}\ }\textbf {\bibinfo
  {volume} {478}},\ \bibinfo {pages} {89} (\bibinfo {year} {2011})},\ \Eprint
  {https://arxiv.org/abs/1106.3614} {arXiv:1106.3614 [quant-ph]} \BibitemShut
  {NoStop}%
\bibitem [{\citenamefont {Chan}\ \emph {et~al.}(2012)\citenamefont {Chan},
  \citenamefont {Safavi-{N}aeini}, \citenamefont {Hill}, \citenamefont
  {Meenehan},\ and\ \citenamefont {Painter}}]{Chan2012}%
  \BibitemOpen
  \bibfield  {author} {\bibinfo {author} {\bibfnamefont {J.}~\bibnamefont
  {Chan}}, \bibinfo {author} {\bibfnamefont {A.~H.}\ \bibnamefont
  {Safavi-{N}aeini}}, \bibinfo {author} {\bibfnamefont {J.~T.}\ \bibnamefont
  {Hill}}, \bibinfo {author} {\bibfnamefont {S.}~\bibnamefont {Meenehan}},\
  and\ \bibinfo {author} {\bibfnamefont {O.}~\bibnamefont {Painter}},\
  }\bibfield  {title} {\bibinfo {title} {Optimized optomechanical crystal
  cavity with acoustic radiation shield},\ }\href
  {https://doi.org/10.1063/1.4747726} {\bibfield  {journal} {\bibinfo
  {journal} {Appl. Phys. Lett.}\ }\textbf {\bibinfo {volume} {101}},\ \bibinfo
  {pages} {081115} (\bibinfo {year} {2012})},\ \Eprint
  {https://arxiv.org/abs/1206.2099} {arXiv:1206.2099 [physics.optics]}
  \BibitemShut {NoStop}%
\bibitem [{\citenamefont {Deotare}\ \emph {et~al.}(2009)\citenamefont
  {Deotare}, \citenamefont {Mc{C}utcheon}, \citenamefont {Frank}, \citenamefont
  {Khan},\ and\ \citenamefont {Lon\v{c}ar}}]{Deotare2009}%
  \BibitemOpen
  \bibfield  {author} {\bibinfo {author} {\bibfnamefont {P.~B.}\ \bibnamefont
  {Deotare}}, \bibinfo {author} {\bibfnamefont {M.~W.}\ \bibnamefont
  {Mc{C}utcheon}}, \bibinfo {author} {\bibfnamefont {I.~W.}\ \bibnamefont
  {Frank}}, \bibinfo {author} {\bibfnamefont {M.}~\bibnamefont {Khan}},\ and\
  \bibinfo {author} {\bibfnamefont {M.}~\bibnamefont {Lon\v{c}ar}},\ }\bibfield
   {title} {\bibinfo {title} {Coupled photonic crystal nanobeam cavities},\
  }\href {https://doi.org/10.1063/1.3176442} {\bibfield  {journal} {\bibinfo
  {journal} {Appl. Phys. Lett.}\ }\textbf {\bibinfo {volume} {95}},\ \bibinfo
  {pages} {031102} (\bibinfo {year} {2009})},\ \Eprint
  {https://arxiv.org/abs/0905.0109} {arXiv:0905.0109 [physics.optics]}
  \BibitemShut {NoStop}%
\end{thebibliography}
%

		
\end{document}